\documentclass[prb,twocolumn]{revtex4}
\usepackage[latin1]{inputenc}
\usepackage{hyperref}
\usepackage{amsmath}
\usepackage{graphicx}
\usepackage{graphics}
\usepackage{amsfonts}
\usepackage{amssymb}
\usepackage{natbib}
\usepackage{subfigure}
\begin{document}
\title{Quantum Oscillations in a $\pi$-Striped Superconductor}
\date{\today}
\author{M. Zelli}
\email{zellim@mcmaster.ca}
\affiliation{Department of Physics and Astronomy, McMaster University, Hamilton ON, Canada}
\author{Catherine Kallin}
\affiliation{Department of Physics and Astronomy, McMaster University, Hamilton ON, Canada}
\author{A. John Berlinsky}
\affiliation{Department of Physics and Astronomy, McMaster University, Hamilton ON, Canada and Perimeter Institute for Theoretical Physics, Waterloo ON, Canada.}
\begin{abstract}
Within Bogoliubov-de Gennes theory, a semiclassical approximation is used to study quantum oscillations and to determine the Fermi surface area associated with these oscillations in a model of a $\pi$-striped superconductor, where the d-wave superconducting order parameter oscillates spatially with period 8 and zero average value.  This system has a non-zero density of particle-hole states at the Fermi energy, which form Landau-like levels in the presence of a magnetic field, B. The Fermi surface is reconstructed via Andreev-Bragg scattering, and the semiclassical motion is along these Fermi surface sections as well as between them via magnetic breakdown. Within the approximation, oscillations periodic in 1/B are found in both the positions and widths of the lowest Landau levels. The area corresponding to these quantum oscillations for intermediate pairing interaction strength is similar to that reported for experimental measurements in the cuprates. A comparison is made of this theory to data for quantum oscillations in the specific heat measured by Riggs et al.
\end{abstract}
\maketitle
\section{Introduction}
The nature of the normal state in the cuprates remains a mystery after decades of research and exploration. There is general agreement that these are strongly correlated systems and considerable evidence for non-Fermi liquid behavior, particularly in the low-doping region of the phase diagram, the so-called pseudogap phase.\cite{pseudo,timusk}.  There is also evidence for competing, broken-symmetry phases, including stripe behavior in the charge and spin density.  The possibility of coexisting or close-by phases suggests that these might be stabilized by the variation of some external parameter, such as pressure or magnetic field.

The observation of quantum oscillations in the electrical resistivity of cuprates in 2007 added one more piece to the puzzle of high temperature cuprate superconductivity.\cite{doiron} Since then, quantum oscillations have been observed in other physical properties and are now a well-established phenomenon in the cuprates.\cite{jaudet,riggs,bangura,yelland,audouard,sebastian,singleton,rourke,yao} The observed quantum oscillations are indicative of a Fermi surface (FS) with an electron pocket\cite{leboeuf} with an area of about $2\%$ of the Brillouin zone (BZ), which is significantly smaller than the area one would expect from band structure calculations. A FS reconstruction approach due to some form of translational symmetry breaking order may explain quantum oscillations and the small area.\cite{densitywave,cdw} However, there are other observations that do not agree with the FS reconstruction approach. One is the ARPES experiments which see only disconnected sections of FS, the so-called Fermi arcs.\cite{arc,fermiarc} Another observation is the specific heat \cite{riggs} which suggests that the $\sqrt{H}$ dependence of the Sommerfeld coefficient persists above the resistive transition. This dependence is associated with d-wave superconductivity. However, its persistence above $\rm T_c$ is surprising.  Furthermore, it was found that the typical FS reconstruction approach produces a specific heat that is too large to be consistent with experimental measurements.\cite{riggs}

In an earlier study,\cite{zelli} we considered the mixed states of a $\pi$-striped superconducting model where a spatially periodic d-wave pairing interaction leads to a reconstructed FS.\cite{stripe} This model has been proposed to explain the 1/8 anomaly which is observed in some of the lanthanum cuprates.\cite{berg}   Surprisingly, we found that, despite particle-hole mixing, Landau levels (LL) - a necessary prerequisite for quantum oscillations - are formed in the low-energy DOS for values of the pairing interaction where the spectral function exhibits Fermi arcs. Additionally, the cyclotron effective mass for this model, defined based on the LL spacing, was shown to be equal to the specific heat effective mass, indicating that FS reconstruction for a $\pi$-striped phase does not necessarily lead to too large a specific heat. Therefore, with the exception of the $\sqrt{H}$ of the background specific heat, which does not occur in this model,  the properties of the $\pi$-striped superconductor that we calculated were consistent with those of cuprates in the presence of a magnetic field. However, our earlier study, which was limited to discrete, well-separated values of magnetic field, did not allow direct calculation of quantum oscillations to obtain an area that could be compared to  experiment. 

In the present study, we employ a semiclassical, approximate method that overcomes the limitations of the previous study and enables us to make more quantitative comparison with experiments. This approximation and a detailed analysis of the behavior as a function of magnetic field, chemical potential and pairing strength, allows us to connect the area associated with quantum oscillations directly to the reconstructed FS of the $\pi$-striped superconductor.  For a physically plausible value of the gap amplitude, the quantum oscillation frequency for the specific heat is found to be close to the experimental value.\cite{riggs}

The remainder of this paper is organized as follows. In Sec.~2 we briefly review the $\pi$-stripe model in zero field and the Fermi surfaces that result for very small and larger gaps. In Sec.~3, we introduce the approximate semiclassical numerical method used to study quantum oscillations. Sec.~4 contains a critical discussion of this semiclassical approximation, comparing the modulated case to previous work on uniform d-wave and also further elucidating the nature of the ``broadened Landau levels" found in our earlier work. 

In Sec.~5 we discuss the semiclassical picture of Pippard for motion of electrons in a magnetic field in the presence of a one-dimensional periodic potential, based on linked orbits in position space, and generalize this picture to the case of a periodic pairing potential.  Section 6 shows the result of this method for very small values of the pairing interaction where the shape of the spectral function at zero energy is close to the unperturbed FS. In Sec.~7, results are shown for larger values of the pairing interaction where the shape of the spectral function resembles Fermi arcs. In addition, some special cases, e.g., half filling and very large gap amplitude, are examined here in order to facilitate identification of the orbits.  Section 8 shows how quantum oscillations in the specific heat behave for this model. Finally, the plausibility and implications of such a superconducting $\pi$-striped model and issues surrounding quantum oscillations in this model are discussed in Sec.~9.

\section{The $\pi$-stripe model in zero field}
The tight-binding mean-field Hamiltonian\cite{shirit} describing a model of a two-dimensional $\pi$-striped superconductor is given by
\begin{align}
H=H_{0}+\sum_{x,y} \Delta \{ \cos(q_{x}x)[c^{\dagger}_{x,y\uparrow}c^{\dagger}_{x+1,y\downarrow}-c^{\dagger}_{x,y\downarrow}c^{\dagger}_{x+1,y\uparrow}] \label{eq:hamil} \\
-\cos(q_{x}(x-1/2))[c^{\dagger}_{x,y\uparrow}c^{\dagger}_{x,y+1\downarrow}-c^{\dagger}_{x,y\downarrow}c^{\dagger}_{x,y+1\uparrow}]+H.C. \} \nonumber
\end{align}
where $c^{\dagger}_{x,y\sigma}$ creates an electron with spin $\sigma$ on site $(x,y)$. $H_{0}$ is the kinetic part of the Hamiltonian with only the first nearest neighbor hopping term, $t$, present. The d-wave-type order parameter has a periodicity of $2\pi/q_{x}$ in the $x$ direction in position space. Here $q_{x}=\pi/4$ corresponds to an $8$-site periodicity of the order parameter.

More details about the model and its dependence on $\Delta$ are provided in Ref. \onlinecite{zelli}. In addition, we make the following observations: (1) In momentum space the periodic pairing potential connects particle states with wavevector $k$ plus an even multiple of $q_{x}$ to hole states with wavevector $-k$ plus an odd multiple of $q_{x}$ and {\it vice versa}.  In other words, these two sets of particle-hole states are decoupled.  This implies that the Bogoliubov-de Gennes Hamiltonian, which is in principle a $16\times 16$ matrix, can be written as two decoupled $8\times 8$ blocks. (2) When plotted in the proper Brillouin zone for this period 8 system, for example $-q_{x}/2 < k \le q_{x}/2$, it is found that the bands stick together at the zone boundary.  This is a result of an additional symmetry of the Hamiltonian under the operation of a translation by 4 lattice spacings, combined with taking $\Delta$ to $-\Delta$.  It is similar to what can happen in the presence of nonsymorphic space group operations.

\begin{figure} [tbph] 
\includegraphics[scale=.55]{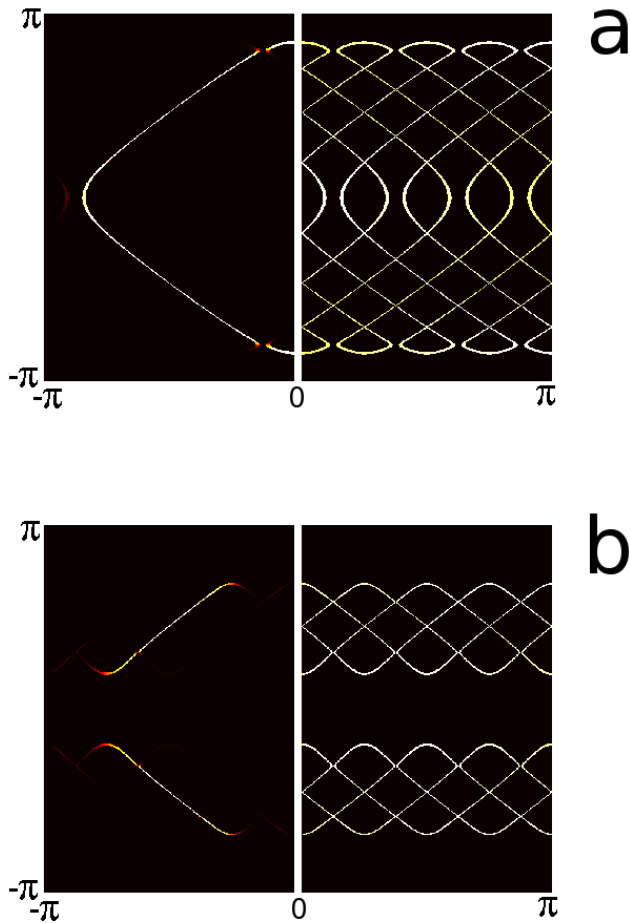}
\caption{One particle spectral functions at zero energy for the period 8 pair density wave system.  The left panels show the spectral weight for adding or removing an electron (or hole) at $E=0$ in the extended BZ while the right panels show the same spectral weight folded into the reduced BZ and then repeated, which illustrates the complex shape of the Fermi surface. All panels are at 1/8 doping for (a) $\Delta=0.02$, and (b) $ \Delta=0.25$.} 
\label{sp02and25}
\end{figure}

In this study, we focus on two ranges of values of $\Delta$. One is the range of very small $\Delta$, $\Delta < 0.05$,  where one can understand the shape of the FS based on a simple perturbative approach. The other is the range of intermediate values of $\Delta$, $0.2 \le \Delta \le 0.3$. The one-electron spectral function at zero energy  for two such cases is shown in Fig.~\ref{sp02and25}.  For $\Delta=0.02$, for the one-particle spectral functions shown in the left hand panel of Fig.~\ref{sp02and25}a, only the small parts of the $\Delta=0$ FS near $(0,\pm\pi)$ that are connected by $\pm q_{x}$ are gapped out. The right hand side of the figure shows the same zero energy, one-electron spectral function folded back into the reduced BZ and then repeated in the $k_{x}$ direction. This locus of non-zero spectral weight at zero energy illustrates the complex Fermi surface of this pair density wave (PDW) system for small $\Delta$.

The shape of the spectral function for $\Delta = 0.25$, shown in the left hand panel of Fig.~\ref{sp02and25}b, is similar to Fermi arcs. The corresponding FS is illustrated in the right hand panel. Other values of $\Delta$ are also discussed in this paper to illustrate the evolution of quantum oscillations from one regime to another.

\section{Semiclassical theory in a field: BdG without vortices}

In our earlier study,\cite{zelli} a magnetic field was incorporated into the model using the so-called Franz-Tesanovic (FT) singular gauge transformation,\cite{dwaveft,fttrans} and the resulting Boboliubov-deGennes (BdG) equations were solved numerically.  A requirement of this approach is that one needs to introduce a bipartite lattice of vortices that are commensurate with the superlattice. As a result, the magnetic field can only be changed in very large steps which makes it impossible to measure the area associated with quantum oscillations. In addition, commensuration effects, due specifically to the assumed perfect order of the vortex lattice, further complicate the analysis. 

In this study, we use a different approach. Consider the $\Delta=0$ case which describes two-dimensional electrons with tight-binding hopping. To apply a magnetic field to the system, one introduces magnetic unit cells. The phase of the hopping term changes by $2\pi$ in going around a magnetic unit cell. In this case, either choice of a square or rectangular unit cell results in the same DOS spectrum for a given magnetic field provided the magnetic unit cells have the same number of sites. Consequently, one can go the limit where the unit cell is a single row of sites. The advantage of using a row unit cell is that one can add only one site to a unit cell to proceed to the next available unit cell size. For a row of length $L$, the fractional decrease in the field for adding one site is $-1/L$. If one uses a wider, shorter magnetic unit cell, say $(L/m) \times m$ (for $L$ a multiple of $m$), then the fractional decrease in field from increasing $L/m$ by one is $-m/L$. The field increments are even larger if one maintains a square aspect ratio. Thus a magnetic unit cell formed by a single line of sites allows field increments of the smallest fractional size. For the rest of this paper, we use $L$ to refer to the number of sites in a magnetic unit cell so that $L=256$ could correspond to a row unit cell of length $256$ or a square unit cell with a linear size $l=16$.

In a superconductor, one can not go to the row limit for a magnetic unit cell because of the supercurrent field associated with the vortex lattice. However, if one assumes that the effect of vortices is negligible, then row unit cells can be used. This enables us to change the magnetic field in much smaller steps and eliminates commensuration effects (which are probably unrealistic for the cuprates), allowing us to study quantum oscillations and determine their frequencies, which can then be associated with orbits in $k$-space. We will refer to the approximation of neglecting the superfluid velocity as the semiclassical approximation or the no-vortex case.

In order to formulate this approximation more explicitly, we consider how vortices enter into the BdG Hamiltonian, starting from the BdG Hamiltonian in a magnetic field. 
\begin{equation}
H=
\begin{pmatrix} 
  -t\displaystyle\sum_{\delta}e^{-iA_{\delta}(r)}\hat{s}_{\delta}-\mu   &  \displaystyle\sum_{\delta}\Delta_{\delta} e^{i\phi(r)/2}\hat{s}_{\delta}e^{i\phi(r)/2} \\  
   \displaystyle\sum_{\delta} \Delta_{\delta} e^{-i\phi(r)/2}\hat{s}_{\delta}e^{-i\phi(r)/2}&    t\displaystyle\sum_{\delta}e^{iA_{\delta}(r)}\hat{s}_{\delta}+\mu
   \end{pmatrix}
\end{equation}
where $\hat{s}_{\delta}$ is defined as the operator, $\hat{s}_{\delta}u(r)=u(r+\delta)$. For a model of a $\pi$-striped superconductor, the space dependent pairing interaction is $\Delta_{\delta}=\Delta \cos(q_{x}(x - 1/2\pm 1/2))$ if $\delta=\pm \hat{x}$ and $\Delta_{\delta}=-\Delta \cos(q_{x}(x-1/2))$ if $\delta=\pm \hat{y}$. $A_{\delta}(r)=\frac{e}{\hbar c}\int_{r}^{r+\delta}A(r)dr$ where $A(r)$ is the vector potential associated with the magnetic field, and $\phi(r)$ is the phase of the order parameter on site $r$.
To eliminate the phase of the order parameter, we apply the following singular gauge transformation
\begin{equation}
U=
\begin{pmatrix} 
  e^{i\phi(r)}    &0 \\   0 &    1
\end{pmatrix}
\end{equation}
which is a single-valued transformation.\cite{anderson} This yields
\begin{equation}
H=
\begin{pmatrix} 
-t\displaystyle\sum_{\delta}e^{-i(A_{\delta}(r)-\nabla \phi_{\delta}(r))}\hat{s}_{\delta}-\mu   &  \displaystyle\sum_{\delta}\Delta_{\delta} e^{i\nabla\phi_{\delta}(r)/2}\hat{s}_{\delta} \\  
 \displaystyle\sum_{\delta} \Delta_{\delta} e^{i\nabla\phi_{\delta}(r)/2}\hat{s}_{\delta}&    t\displaystyle\sum_{\delta}e^{iA_{\delta}(r)}\hat{s}_{\delta}+\mu
   \end{pmatrix}
\end{equation}
where $\nabla \phi_{\delta}(r)=\phi(r+\delta)-\phi(r)$. Now using the definition of the superfluid velocity, we can write the Hamiltonian as follows
\begin{equation}
\begin{pmatrix} 
-t\displaystyle\sum_{\delta}e^{i(A_{\delta}(r)+2v_{s}^{\delta}(r))}\hat{s}_{\delta}-\mu   &  \displaystyle\sum_{\delta}\Delta_{\delta} e^{i(A_{\delta}(r)+v_{s}^{\delta}(r))}\hat{s}_{\delta} \\  
 \displaystyle\sum_{\delta} \Delta_{\delta} e^{i(A_{\delta}(r)+v_{s}^{\delta}(r))}\hat{s}_{\delta}&    t\displaystyle\sum_{\delta}e^{iA_{\delta}(r)}\hat{s}_{\delta}+\mu
   \end{pmatrix}
\label{HBdG}
\end{equation}
where $mv_{s}^{\delta}(r)=\hbar \nabla \phi_{\delta}(r)/2-e/c A_{\delta}(r)$.

We note that, although Eq.~(\ref{HBdG}) was derived using the Anderson gauge, essentially the same result can easily be derived in the FT gauge because the superfluid velocity is gauge invariant.  The only difference is that, for the FT gauge, the vector potential, $A_{\delta}(r)$ in Eq.~\ref{HBdG}, is replaced by, $-v_{\delta}^B(r)$, the superfluid velocity field of the $B$ vortices, which, in the usual form of the FT Hamiltonian, acts only on the holes. 

If the effect of vortices is negligible, one can set $v_{s}^{\delta}(r)=0$ in the BdG Hamiltonian and work with a row magnetic unit cell. In the following sections, we apply this approximation and compare the results to that of the full BdG equations with vortices to check whether the approximation is useful.

The length of a row unit cell, which is spanned in the $x$ direction, is given by $L=8m$ where $m$ is an integer. The magnetic field associated with a unit cell $L$ lattice constants long is $B=\phi_{0}/La^{2}$ where $a$ is the lattice spacing. The number of magnetic unit cells in the $x$ direction can be taken to be only one because adding more unit cells in the $x$ direction results in the same DOS spectrum. However, the number of unit cells in the $y$ direction, $N$, must be large to give a well-defined DOS. Using Bloch's theorem, one needs to diagonalize $N$ BdG  matrices with linear size $2L$ so that the total number of positive-energy states is $NL$.  These properties of the spectra and the broadened Landau levels that result are discussed in more detail in the next section.

\section{Critique of the Semiclassical Approximation}
The electronic states of a d-wave superconductor in the presence of a perpendicular magnetic field have been the subject of intense theoretical scrutiny as well as some controversy.  Early on, it was suggested by Gor'kov and Schrieffer\cite{gorkov} and by Anderson\cite{anderson} that the spectrum in a magnetic field consisted of Landau levels with energies $\pm\hbar\omega_H\sqrt{n}$ where $n$ is a positive integer and $\omega_H = \sqrt{2\omega_c\Delta/\hbar}$, $\omega_c$ is the cyclotron frequency, and $\Delta$ is the maximum gap\cite{marinelli},  which is essentially the result for an anisotropic Dirac cone.  A key step in obtaining this result is the neglect of the superfluid velocity due to vortices compared to the vector potential {\bf A}.  However, it was soon shown by Mel'nikov\cite{melnikov} that the superfluid velocity is a strong perturbation for this problem, and, not long afterward, Franz and Tesanovic\cite{fttrans} developed a solution  which treated the superfluid velocity field and the vector potential on an equal footing.  These were expressed in terms of two superfluid velocity fields, one interacting with electrons and the other with holes.  The problem reduces to that of Dirac quasiparticles in the presence of an effective scalar potential and a vector potential corresponding to zero average magnetic field, and the solutions can be expressed in terms of Bloch functions.  The resulting quasiparticle spectra are gapless and band-like, although Landau levels appear at high energies, well above the scale of the gap.\cite{melnikov,dwaveft}.

Vafek et al.\cite{dwaveft} give an elegant expression for the continuum Hamiltonian near a single node in the FT gauge.
\begin{equation}
H_{N} = v_{F}(p_{x} + a_{x})\tau_{3} + v_{\Delta} (p_{y} + a_{y})\tau_{1} +mv_{F} v_{sx},
\label{dirac}
\end{equation}
where $\tau_i$ are Pauli matrices, $2{\bf a} = m({\bf v}_s^A - {\bf v}_s^B)$, and $2{\bf v}_s = {\bf v}_s^A + {\bf v}_s^B$. This expression shows that the difference of the superfluid velocities associated with the vortices on the two sublattices in the FT gauge, the so-called Berry gauge field, acts like a vector potential coupled minimally to the momenta of the Dirac quasiparticles, while the  superfluid velocity enters as a scalar potential.  This symmetry of the Hamiltonian results in the spectrum of d-wave quasiparticles in a magnetic field remaining gapless.

We have reviewed the properties of a uniform d-wave superconductor in a magnetic field in such detail in order to emphasize the {\em differences} between that problem and the one treated in this paper and in our earlier work.\cite{zelli} The main difference is that the $\pi$-striped superconductor in zero field does not have nodal points at $E_F$ with Dirac-like excitations. Instead it has extended regions of Fermi surface which persist in the presence of an off-diagonal potential that couples electron states at $\bf k$ and hole states at ${-\bf k} \pm {\bf q}$ and which gaps only parts of the Fermi surface. Furthermore, for not very large pairing gap amplitude, it is as if $v_{\Delta}$ in Eq. (\ref{dirac}) is essentially zero over most of the Fermi arcs. We also find that, for these gap amplitudes, the $u$'s and the $v$'s, the Bogoliubov quasiparticle amplitudes, jump sharply on crossing the FS (see Fig. 6 of Baruch and Orgad\cite{shirit}), as they do for a normal metal, rather than varying smoothly as they do away from a Dirac point.  These features of the $\pi$-striped superconductor lead to very different behavior in a magnetic field from that of a uniform d-wave superconductor.

Our earlier work demonstrated that the low energy excitation spectra of the $\pi$-stripe phase in a magnetic field consists of broadened but well defined Landau-like levels, with spacings that are linear in magnetic field, for certain ranges of gap amplitude.\cite{zelli} 
Specifically we found that weakly perturbed Landau levels are observed for $0 < \Delta \lesssim 0.07$.  In this range an electron LL is accompanied by a small reflected hole peak and vice versa.  For $0.07 \lesssim \Delta \lesssim 0.13$ and 1/8 doping, Landau levels are not observed, possibly because they are broadened or closely spaced because of sharp structure in the low energy density of states for this range of $\Delta$. (See Fig.~2 of Zelli et al.\cite{zelli})  For $0.14 \lesssim  \Delta \lesssim 0.3$, Landau levels reappear, but, in this region, their amplitude is approximately equal, above and below $ E = 0$. Distinct Landau levels are not observed for gap values larger than $\Delta \approx 0.35$, where the Fermi surface resembles disconnected asymmetrical figure eights.

The above discussion applies to the full BdG equations for a $\pi$-striped superconductor, assuming a square arrangement of vortices positioned on the gap nodes, although the equilibrium arrangement of vortices (and consequently the equilibrium superfluid velocities) for this system has not yet been determined.  What can we say about the validity of the approximation of neglecting the superfluid velocity in Eq.~(\ref{HBdG})? First we note that the semiclassical calculation gives broadened Landau levels for all non-zero values of the modulated gap amplitude, unlike the full BdG equations. It also gives sharp Landau levels in the limit when the gap goes to zero which connect continuously to the weakly broadened levels for small gap. One measure of the domain of validity of the semiclassical approximation is how well the broadened Landau level spectra agree with the density of states for the exact calculation when the superfluid velocity is included. 
As we shall see below, for small $\Delta$ the semiclassical result does a good job of modeling weakly perturbed Landau levels.  Furthermore, for the range of larger values of $\Delta$ that we expect to be relevant to the cuprates, $0.14 \lesssim \Delta \lesssim 0.3$, the semiclassical approximation also agrees well with the exact spectrum (except for lattice commensurability effects near $E = 0$ which will be discussed below).  Since quantum oscillations arise from the presence of broadened Landau levels in the semiclassical approximation, we expect this approximation to be valid in parameter ranges where the exact calculation also exhibits Landau levels.  Conversely, if broadened Landau levels are not present in the full BdG calculation, then the semiclassical approximation is not applicable.

\begin{figure} [tbph] 
\includegraphics[scale=.37]{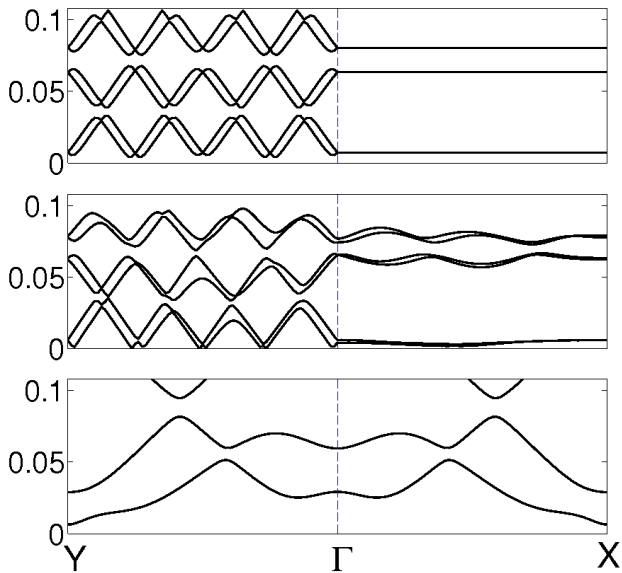}
\caption{Low-lying energy levels for, from top to bottom, modulated d-wave ignoring the effects of vortices, modulated d-wave including the effects of vortices, and uniform d-wave including the effects of vortices.  Here $\Delta = 0.25$, $\mu = -0.3$, and $L = 256$.}
\label{bands}
\end{figure}
To further test the validity of the above arguments, we have calculated the band structure of the excitations for the system with and without vortices for the case $\Delta = 0.25$ and $\mu = -0.3$, using the FT gauge, for a magnetic field corresponding to $L = 256$, along the directions $Y \rightarrow \Gamma \rightarrow X$ in the magnetic BZ. For the semiclassical approximation, Fig.~\ref{bands}a, the bands are flat along $\Gamma \rightarrow X$ due the the symmetry of the magnetic translation group.  Dispersion arises along the $Y$ direction, resulting in one-dimensional density of states peaks. To understand the dispersion along $k_y$ for $v_s=0$ we refer to the continuum picture where, in Landau gauge, the wave functions are plane waves of wavevector $k_y$ along $y$ and localized in $x$ around a position $x_0 \sim k_y$.  In the presence of a spatially varying potential and a perpendicular magnetic field, electrons move along equipotentials.  For a potential modulated along $x$, this motion is along $y$, at a position and energy that depend on $k_y$.  This provides an interpretation of the meaning of ``broadened Landau levels."  The broadening is due to the different ways that a Landau level wave function averages over the periodic (pairing) potential, depending on its position with respect to the modulation.  Comparing Figs.~\ref{bands}a (without vortices) and \ref{bands}b (with vortices), we see that band structure is relatively insensitive to the  effect of vortices for these parameters. Nevertheless it is clear that the broadened Landau levels are further broadened and shifted by the vortices.  To complete this analysis, we also show, in Fig.\ref{bands}c, the energy bands for a uniform d-wave gap of the same magnitude, with vortices and with the same chemical potential and magnetic field.  There is no hint of broadened Landau levels for this case. Although not shown in Fig.\ref{bands}, the semiclassical result for uniform d-wave, neglecting $v_s$, is perfectly flat bands, at energies scaling approximately as $\pm E_1\sqrt{n}$ for $n= 0, 1, 2, \dots$.  For the parameters of Fig.~\ref{bands}c, $E_1\approx 0.31$,  confirming that $v_s$ is indeed a strong perturbation for uniform d-wave.
\begin{figure} [tbph] 
\includegraphics[scale=.26]{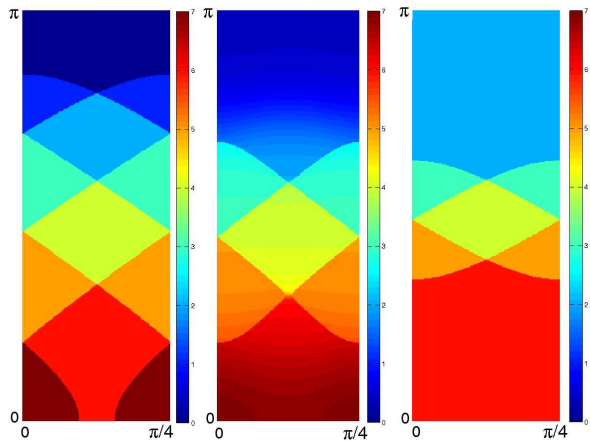}
\caption{The occupation function in the reduced BZ for 3 cases from left to right: a) $\Delta=0$, $V=0$ and $\mu=-0.23$ b)  $\Delta=0.25$, $V=0$ and $\mu=-0.3$ c)   $\Delta=0$, $V=0.7$ and $\mu=-0.05$. $V$ is the magnitude of the interactions in CDW with periodicity of 4 sites. The reduced BZ for the CDW spans from 0 to $\pi/2$. Here, however, we have folded its FS once for easier comparison to the $\pi$-striped superconductor.} 
\label{3nk}
\end{figure}

\begin{figure} [tbph] 
\includegraphics[scale=.29]{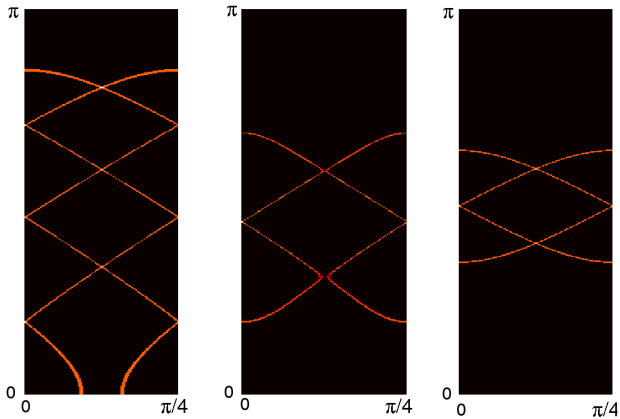}
\caption{FS for 3 cases from left to right: a) $\Delta=0$, $V=0$ and $\mu=-0.23$ b)  $\Delta=0.25$, $V=0$ and $\mu=-0.3$ c)   $\Delta=0$, $V=0.7$ and $\mu=-0.05$. $V$ is the magnitude of the interactions in CDW with periodicity of 4 sites. The reduced BZ for the CDW spans from 0 to $\pi/2$. Here, however, we have folded its FS once for easier comparison to the $\pi$-striped superconductor.} 
\label{3fs}
\end{figure}

One might wonder why LLs should exist in a superconductor, particularly close to $E_F$ where electrons and holes can mix and quasiparticles need not have a definite charge. First, we note that these LLs are different from the sharp LLs of the normal state or Dirac-like LLs in graphene. In our earlier work \cite{zelli}, we showed that there is a periodic structure of peaks in the low-energy DOS, with spacing proportional to $B$, where the sum of the number of states under the two peaks related by reflection about $E=0$ is equal to two times the degeneracy of one LL. We called these peaks Landau levels. Second LLs are not observed for $\Delta > 0.35$, where there is noticeable particle-hole mixing on the Fermi surface, i.e., where the $u$'s and $v$'s differ noticeably from 0 or 1 on either side of the Fermi surface.  LLs are only observed for values of $\Delta$ where the quasiparticle charge close to the FS is effectively quantized as either $e$ or $-e$. This is illustrated in Fig.~\ref{3nk} where the electron occupation number, $n(k)$, is shown in the reduced Brillouin zone for the case of $\Delta = 0.25$ and compared to $n(k)$ in the normal metal, $\Delta = 0$,  as well as the case of a charge density wave (CDW). The CDW parameters have been chosen to give a FS which closely resembles that of the PDW at $\Delta = 0.25$.  The Fermi surfaces for these three cases are shown in Fig.~\ref{3fs}.

\begin{figure} [tbph] 
\includegraphics[scale=.2]{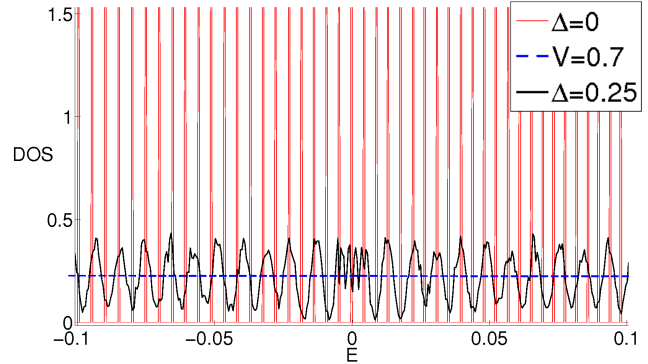}
\caption{Low-energy DOS in the presence of a magnetic field of $L=1024$ for 3 cases: a) $\Delta=0$, $V=0$ and $\mu=-0.23$ b)  $\Delta=0.25$, $V=0$ and $\mu=-0.3$ c)   $\Delta=0$, $V=0.7$ and $\mu=-0.05$.} 
\label{lowfield}
\end{figure}

For intermediate $\Delta$, (as in Fig.~\ref{sp02and25}b) the parts of the FS that are affected to first-order in the pairing potential term are gapped out. At the edge of the Fermi arcs, the rounded Fermi surface sections are controlled by $\Delta^{2}$ terms which act like a normal potential. The small gaps within Fermi arcs are due to $\Delta^{3}$ terms which cause particle-hole mixing. In Figs.~\ref{3fs}b and \ref{3fs}c, we compare the FS of a $\pi$-striped superconductor (PDW) with $\Delta=0.25$ to that of a CDW with $V=0.7$. The small $\Delta^3$ gaps can be seen in  the center of Fig.~\ref{3fs}b and are absent in Fig.~\ref{3fs}c.  Since, even for intermediate $\Delta$ (e.g. $\Delta = 0.25$), $\Delta^3$ is very small, the region with particle-hole mixing in Fig.~\ref{3fs}b is  also small.  Consequently the Fermi surface topology and $n(k)$ near the Fermi surface are almost the same for the cases of a PDW and a CDW in Figs.~\ref{3nk} and \ref{3fs}.  However, as shown in Fig.~\ref{lowfield}, the PDW supports broadened Ls whereas the CDW does not. No LLs are expected for the CDW case as the Fermi surface consists of open orbits.  These results imply that the $\Delta^3$ Andreev-scattering is essential to the formation of broadened LLs in the PDW case.  Since the $\Delta^3$ Andreev scattering gives rise to closed orbits, albeit closed orbits which are partly electron-like and partly hole-like, it seems natural to consider the possibility that these closed orbits are responsible for the broadened LLs.\cite{footnote,pereg-barnea}
If so, do they also support quantum oscillations with a frequency related to their area?  This is the question that our semiclassical approximation addresses.

\section{Pippard's Semiclassical Picture}
It will be useful for understanding quantum oscillations for the $\pi$-striped superconductor in a magnetic field to first consider a more traditional semiclassical picture of the effect of a magnetic field on the motion of electrons in a 2D layer. For simplicity we start with a circular FS.  The presence of a weak periodic potential causes gaps in the FS segments which reconstruct in the reduced BZ, leading to more complicated orbital motions. This will be the case both for periodic potentials and for periodic pairing potentials. The analysis is particularly straightforward for the case of weak periodic potentials.

\begin{figure} [tbph] 
\includegraphics[scale=.27]{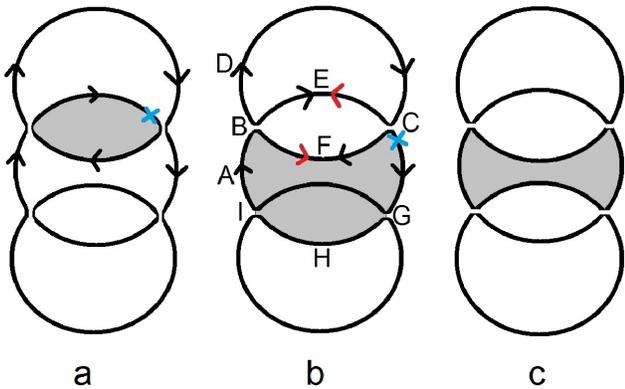}
\caption{Semiclassical motion of a nearly free particle system in the presence of a weak periodic potential (a) and a weak periodic superconducting pairing potential, (b and c). The direction of the semiclassical motion for particles is shown by black arrows. Holes (shown by red arrows) precess in the opposite direction. The gray area in the center figure is $(\frac{\hbar c}{eH})^{2}(A_{T}-A_{b})$ where $A_{b}$ is the area of the small electron pocket in panel a and $A_{T}$ is the area of the original circular FS. Starting from the blue cross in panel b, the particle can either go over the whole unperturbed circular orbit by tunnelling at points G,I,B, and C, or tunnel only at points G and I and Andreev scatter twice at points B and C covering the gray area. Another possible path is to Andreev scatter at points G and I and tunnel at points B and C. However, this path covers the same gray area.  The change in the phase of the wave function is $\frac{\hbar cA_{T}}{eH}$ when the particle goes over the whole circular circuit and $\frac{\hbar c(A_{T}-A_{b})}{eH}+\beta$ when it travels around the shaded area, where $\beta$ is the phase shift due to two consecutive Andreev scatterings and is assumed to be relatively field independent. This behavior should be contrasted to that of the linked orbit of Pippard, shown on the left, where the particles orbit around the areas $A_T$ and $A_b$. Thus, as discussed in the text, the areas associated with quantum oscillations in the width of the first LL are different for the periodic potential and the periodic pairing models. Panel c shows the closed orbit corresponding to four successive Andreev-Bragg scatterings.} 
\label{link}
\end{figure}

To understand these motions, we follow a simple picture due to Pippard.\cite{pippard1962} Pippard introduced the concept of linked orbits where a network of coupled orbits in position space is used to provide a simple and plausible picture of the perturbation of circular electron orbits. This is pictured in Fig.~\ref{link}a, where circular orbits are displaced by the spatial period of the potential. Due to the periodic potential, particles can Bragg scatter from one orbit to another. This results in electron pockets, such as the shaded region, where electrons Bragg scatter twice going around an orbit, with open orbits on either side. For free electrons, the trajectory in $k$ space has the same form as the trajectory in real space, rotated by $\pi/2$. The shaded area in Fig.~\ref{link}a is $(\frac{\hbar c}{eH})^{2}A_{b}$ corresponding to a small electron pocket where, in Pippard's notation, $A_{b}$ is the corresponding area in $k$ space.
For weak periodic potentials and strong magnetic fields, tunneling through the gaps (magnetic breakdown) is highly probable, and the electron motion can also follow the original circular orbit with $k$-space area $A_{T}$ in Pippards notation.

Next we consider what happens for a weak periodic superconducting pairing potential, for which the possible orbits are shown in Fig.~\ref{link}b. Again, for the case of a weak pairing potential and a strong magnetic field, it is possible for electrons or holes to tunnel through gaps at points B, C, G, and I, following the original cyclotron orbit.  For the simplest process involving the periodic pairing potential, a particle could start at the blue X below point C, tunnel at points G and I through section H, and Andreev scatter into a hole at point B, pass point F and Andreev scatter back into a particle at point C. In the first case, the increment in the phase of the wave function is $\frac{\hbar c A_{T}}{eH}$, corresponding to the original FS area.  In the second case it is $\frac{\hbar c(A_{T}-A_{b})}{eH}+\beta$, where $\beta$ is a phase shift due to two consecutive Andreev scatterings and is assumed to be relatively field independent. Note that it is equally possible for the particle to Andreev scatter at points G and I and tunnel at points B and C, and this path covers the same area as in the second case.~\cite{pippard1962}  The probability of undergoing 4 consecutive Andreev reflections (at points B, C, G, and I), corresponding to an area $A_T-2A_b$ and shown in panel c, is small for small $\Delta$ as is discussed further below.  

For fixed chemical potential, $A_{T}$ and $A_{b}$ are fixed. As a result, the phase of the wave function due to different trajectories changes as $H$ is varied. The relative change of the phase due to the two trajectories described above is $\delta \phi=\frac{\hbar A_{b}}{eH}-\beta$. The broadening of a LL will be minimal when $\delta \phi$ is an integral multiple of $2\pi$. The frequency of this occurring and hence the broadening of the LL is then proportional to $A_{b}$ as the magnetic field varies. We will demonstrate below that this is what happens for a striped superconductor. Note that the argument above is not dependent on the symmetry of the order parameter. In fact, for an oscillating s-wave order parameter, the frequency of broadening corresponds to the same area.
We also note that this relationship between the phase along the trajectories and the magnetic field could break down when the effects of superfluid flow are included.\cite{palee-pc}

In his original work, Pippard applied this argument to the broadening of LLs for a real periodic potential.  In this case, it is the interference between the phase shift around the small electron pocket in Fig.~\ref{link}a and that of the original FS that leads to broadening of Landau levels, and the relative phase is $\delta \phi=\frac{\hbar c(A_{T}-A_{b})}{eH}-\beta^{\prime}$. where $\beta^{\prime}$ is the phase shift due to two consecutive Bragg scatterings. We have confirmed through numerical calculations for our model in the normal state with a period 8 site potential, that $A_{T}-A_{b}$ is the area associated with the oscillations in the width of the first LL.

\section{Results for Small $\Delta$}

\begin{figure} [tbph]
\includegraphics[scale=.2]{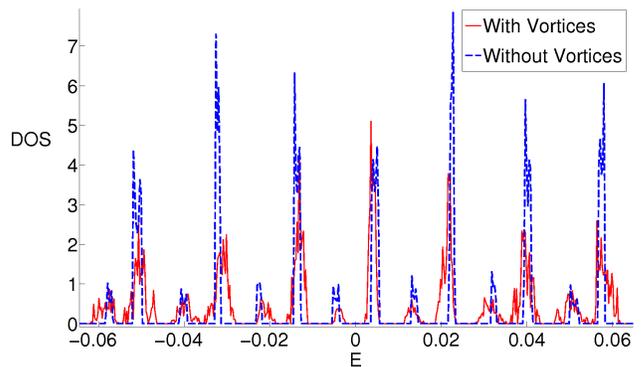}
\caption{Comparison of the low-energy DOS of a $\pi$-striped superconductor with $\Delta=0.02$ and $\mu=-0.23$ in the presence of a magnetic field of $L=256$ with and without vortices, as described in the text.} 
\label{dossmall}
\end{figure}

\begin{figure} [tbph]
\includegraphics[scale=.25]{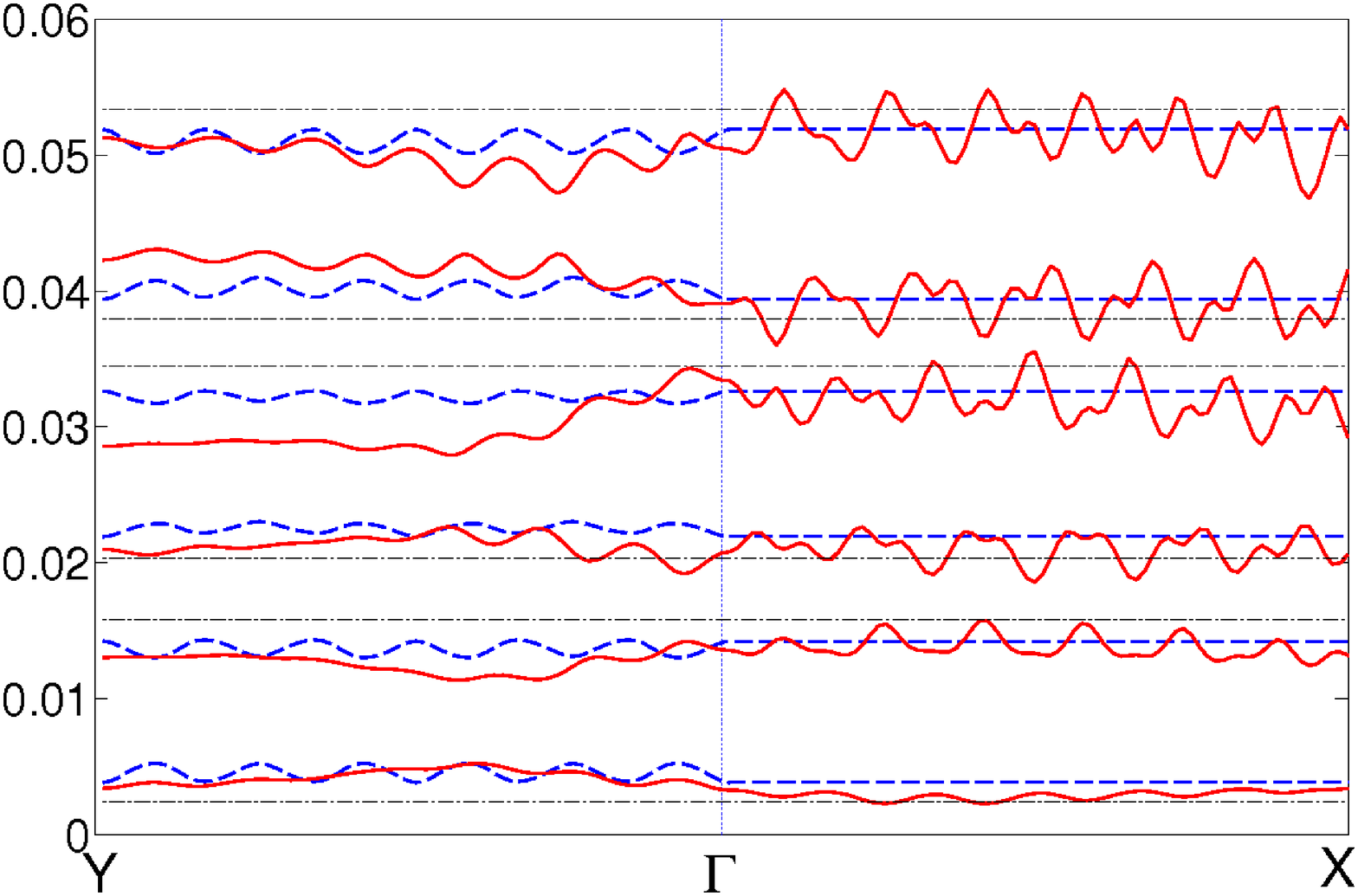}
\caption{Comparison of the low-energy bands for the BdG Hamiltonian of a $\pi$-striped superconductor with $\Delta=0.02$ and $\mu=-0.23$ in the presence of a magnetic field of $L=256$. The solid curves are the bands for the full BdG Hamiltonian, including vortices.  The dashed curves are the semiclassical results for no vortices, and the flat lines (dash-dot lines) are the Landau levels in the limit $\Delta = 0$.} 
\label{smalldelta}
\end{figure}

For small values of $\Delta$, the effect of the pairing interaction is to induce small gaps in the closed $\Delta=0$ FS as shown in Fig.~\ref{sp02and25}a. For these values, numerical results with and without vortices, result in similar low-energy DOS as shown in Fig.~\ref{dossmall} for $\Delta=0.02$ and $L=256$ at $1/8$ doping. The energy bands, shown for positive energy along $Y\rightarrow \Gamma \rightarrow X$ in Fig.~\ref{smalldelta}, look similar for the cases with and without vortices and both are similar to the Landau level structure observed for $\Delta = 0$. Note that the bands for both cases, with and without vortices, shift in the same direction from the $\Delta = 0$ Landau level energies.

\begin{figure} [tbph] \label{broadsmall}
\includegraphics[scale=.21]{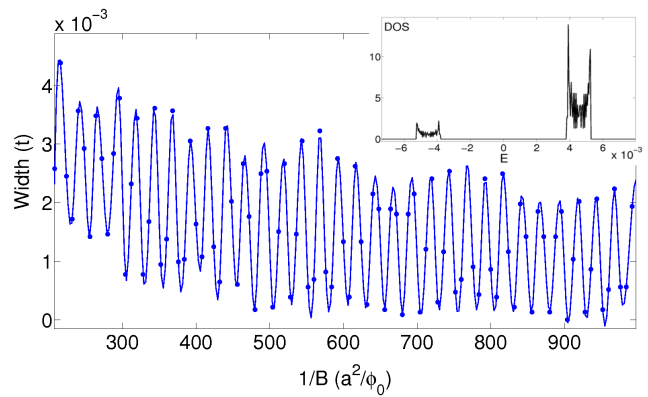}
\caption{The width of the LL closest to $E=0$ as a function of $1/B$ or $L$ for $\Delta=0.02$ and $\mu=-0.23$, corresponding to $1/8$ doping. $1/B$ is written in terms of the lattice constant, $a$, and flux quantum, $\phi_{0}$. The solid line is a spline fit to the data that shows the oscillatory behavior more clearly. The inset shows the first LL for $L=256$.} 
\label{emu23}
\end{figure}

At first, it seems that the only effect of the small pairing potential is to partially reflect each unperturbed LL to the other side of the Fermi energy. This suggests that the area associated with quantum oscillations should remain the large closed FS area for $\Delta=0$. However, as we have seen, this is not the whole story.  It also happens that interference between the original FS area and another orbit induced by the potential leads to LL broadening which oscillates as a function of magnetic field. The widths of the Landau levels near the Fermi energy affect the low temperature properties of the system and, consequently, their dependence on magnetic field is expected to be experimentally observable. In the discussion which follows, we focus on the LL closest to the Fermi energy and measure its width and its position relative to the Fermi energy. We refer to this LL as the first LL. Here, we define the width to be the difference between the low and high energy ends of a LL feature in the DOS spectrum (see the inset of Fig.~\ref{emu23}). By choosing a large system size $N$ in the $y$ direction and sufficiently small energy intervals for the DOS calculation, the width of a LL can be calculated with precision.

The width of the first LL as a function of $1/B$ is shown in Fig. \ref{emu23} for $\Delta=0.02$ and $\mu=-0.23$, corresponding to 1/8 doping. As expected, the width shows an oscillatory behavior. We have argued that the frequency of these oscillations should be related to the differences in areas of FS orbits.
\begin{figure} [tbph] 
\includegraphics[scale=.21]{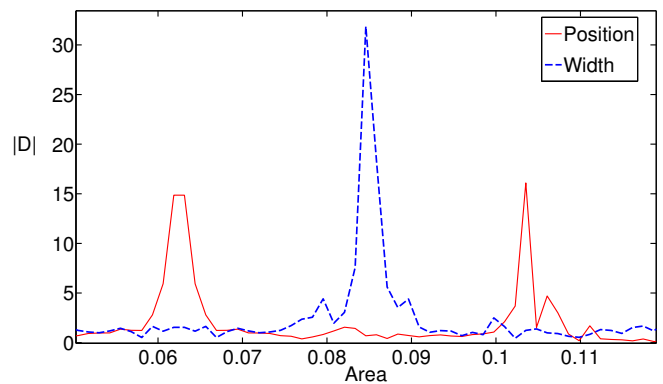}
\caption{Power spectrum associated with the oscillations of the width and position of the first LL for $\Delta=0.02$ at $1/8$ doping. The $x$ axis is rescaled so that it corresponds to area in units of the area of BZ.} 
\label{power}
\end{figure}

Figure~\ref{power} shows the power spectrum associated with oscillations in the width of the first LL for $\Delta=0.02$ and $\mu=-0.23$. The $x$ axis has been rescaled to correspond to area in units of the area of the BZ. The peak in the power spectrum associated with oscillations in the width occurs at an area of about 0.0845.  Note that, since a minimum of $8$ sites must be added to a magnetic unit cell in changing $B$, one can not directly measure periods of oscillations in $L \propto 1/B$ that are smaller than $8$. This means that the area measured by the power spectrum analysis is, in fact, an area modulo $1/8$.  

\begin{figure} [tbph] 
\includegraphics[scale=.15]{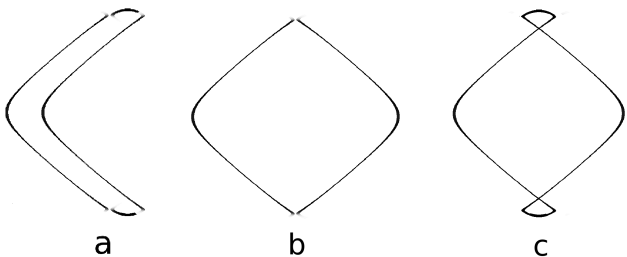}
\caption{(a) Boomerang-shaped FS orbit involving two Andreev-Bragg scatterings and two tunnellings, as shown schematically in Fig.~\ref{link}b, but for a period 8 modulation. The area of this orbit is denoted $A_{T}-A_{b}$ in the text. (b) The corresponding area $A_{b}$.  (c) The area $2A_b-A_T$, corresponding to the difference of figures (a) and (b).} 
\label{boomerang}
\end{figure}

\begin{figure} [tbph] 
\includegraphics[scale=.19]{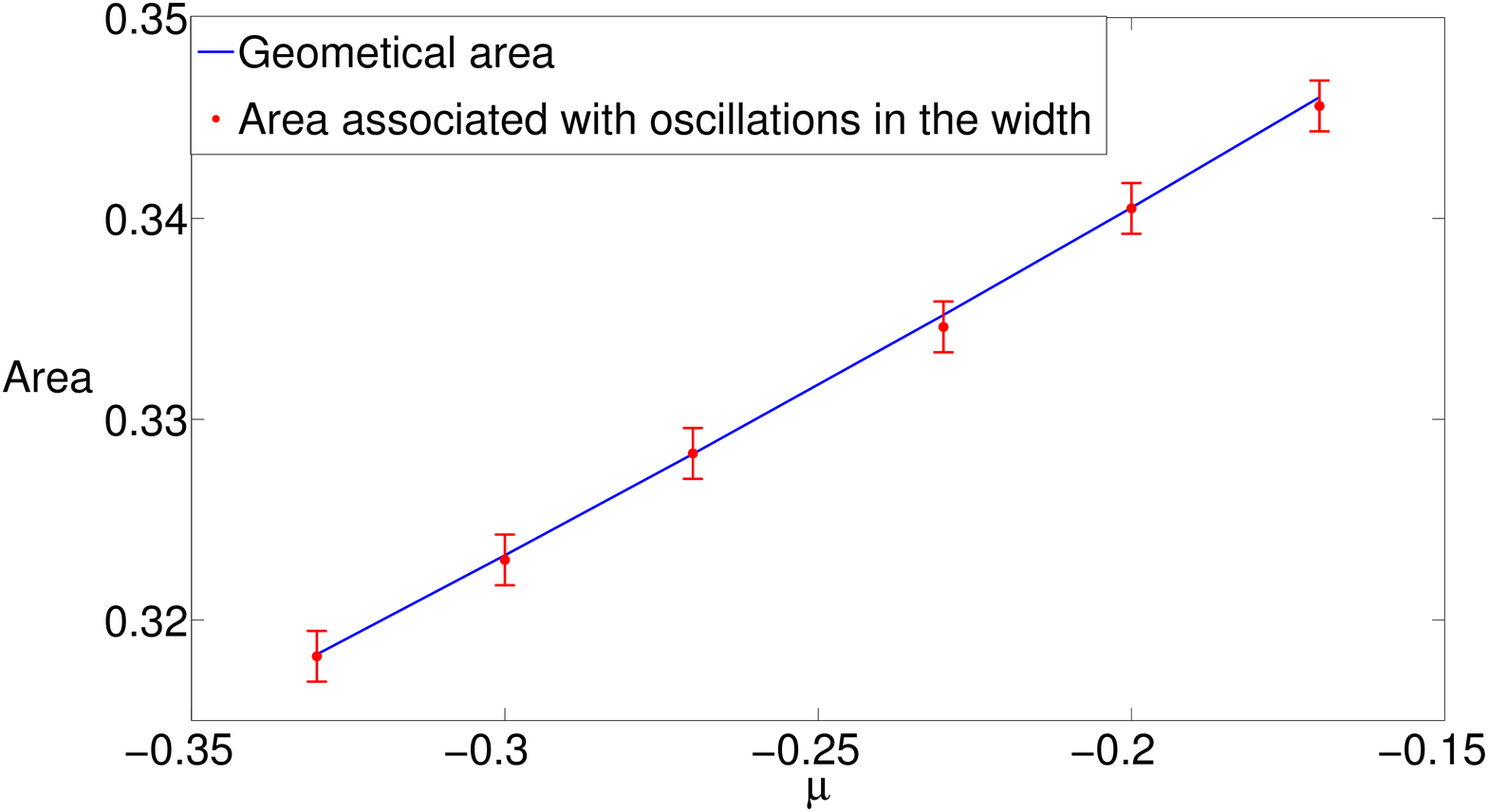}
\caption{Comparison of the geometrical area, $A_b$, and the area associated with quantum oscillations in the width of the first LL for $\Delta=0.02$ vs $\mu$ in the region around $1/8$ doping.} 
\label{areavsmu}
\end{figure}

What do we expect for the FS area corresponding to the frequency of oscillation in the width of the lowest lying LL for this period 8 system?  Comparing Fig.~\ref{sp02and25}a and Fig.~\ref{link}b, we look for the FS trajectory in the former that involves two Andreev-Bragg scatterings and two places where tunnelling occurs across a gap. This orbit has the boomerang-like shape shown in Fig.~\ref{boomerang}a and corresponds to the FS area that we have called $A_{T}-A_{b}$.  Then the area $A_{b}$ is the difference between that of the boomerang and that of the original FS as shown in Fig.~\ref{boomerang}b. This latter area is considerably larger than the value 0.0845 found in the power spectrum of  the width in Fig.~\ref{power}.  Its value is equal to $0.0845 + 2/8 = 0.3345$. To confirm the relation between oscillations in the width versus $1/B$ and the area $A_b$, we measure these oscillations for different values of $\mu$ and see that they track the variation of $A_b$ with $\mu$ as shown in Fig.~\ref{areavsmu}. All of the data points in Fig.~\ref{areavsmu} were obtained by adding $2/8$ to the position of the peak in the power spectrum of oscillations in the LL width.

\begin{figure} [tbph] 
\includegraphics[scale=.19]{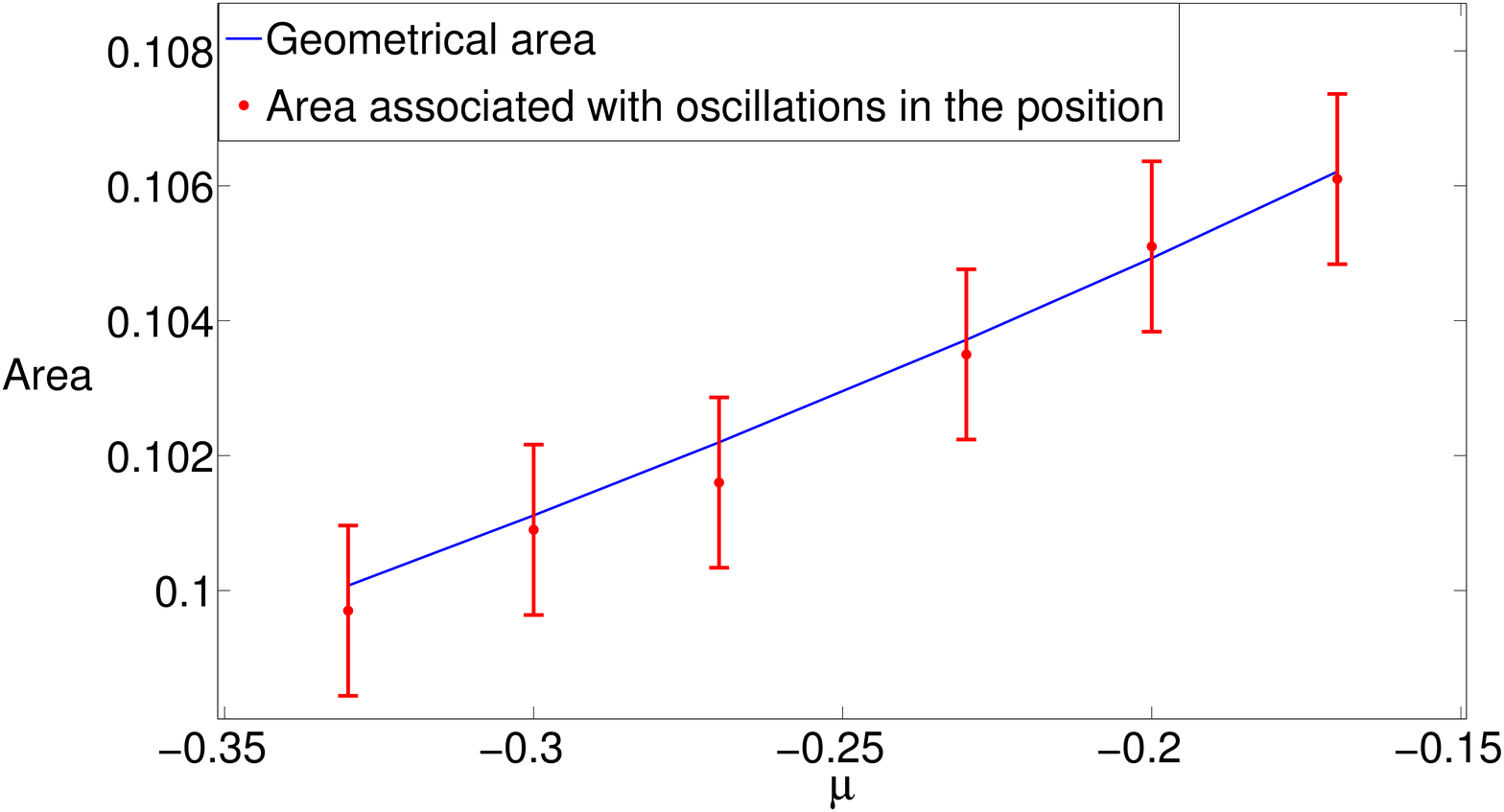}
\caption{Comparison of the geometrical area, $A_{T}-A_{b}$, the boomerang-shaped area in Fig.~\ref{boomerang}a, and the area associated with oscillations in the position of the first LL, corresponding to the highest frequency peak in Fig.~\ref{power}, shown as a function of $\mu$.} 
\label{secpeak}
\end{figure}

Next we consider oscillations in the position of the first LL. Since, the shape of a LL is not symmetric around its position, we define the position of a LL to be the energy at which there are equal numbers of states on both sides. Interestingly, we find that two peaks appear in the power spectrum of the position as shown in Fig.~\ref{power}. The peak on the left corresponds to the $\Delta=0$ FS area, $A_T$, as expected. For this case one must add $3/8$ to the measured value to obtain the actual value of the area. The relevant peak in Fig.~\ref{power} occurs at around $0.0625$ which gives $0.0625 + 3/8 = 0.4375$ for the area of the original FS, corresponding to a density of $0.4375*2=0.875$ electrons per site, as expected for 1/8 doping. The other peak of the power spectrum of position oscillations is associated with $A_{T}-A_{b}$, the  area of the boomerang. From the determinations of $A_b$ and $A_T$ given above, one expects this peak to occur at $0.4375 - 0.3345 = 0.103$ in agreement with the position of the right hand peak in Fig.~\ref{power}. 
The relationship is also confirmed in Fig. \ref{secpeak} where the position in the power spectrum and the geometrical value of $A_{T}-A_{b}$ are compared as $\mu$ is varied.
The picture that emerges is one in which the particles spend part of the time orbiting the original FS and part going around the boomerang-shaped surface.

\begin{figure} [tbph] 
\includegraphics[scale=.2]{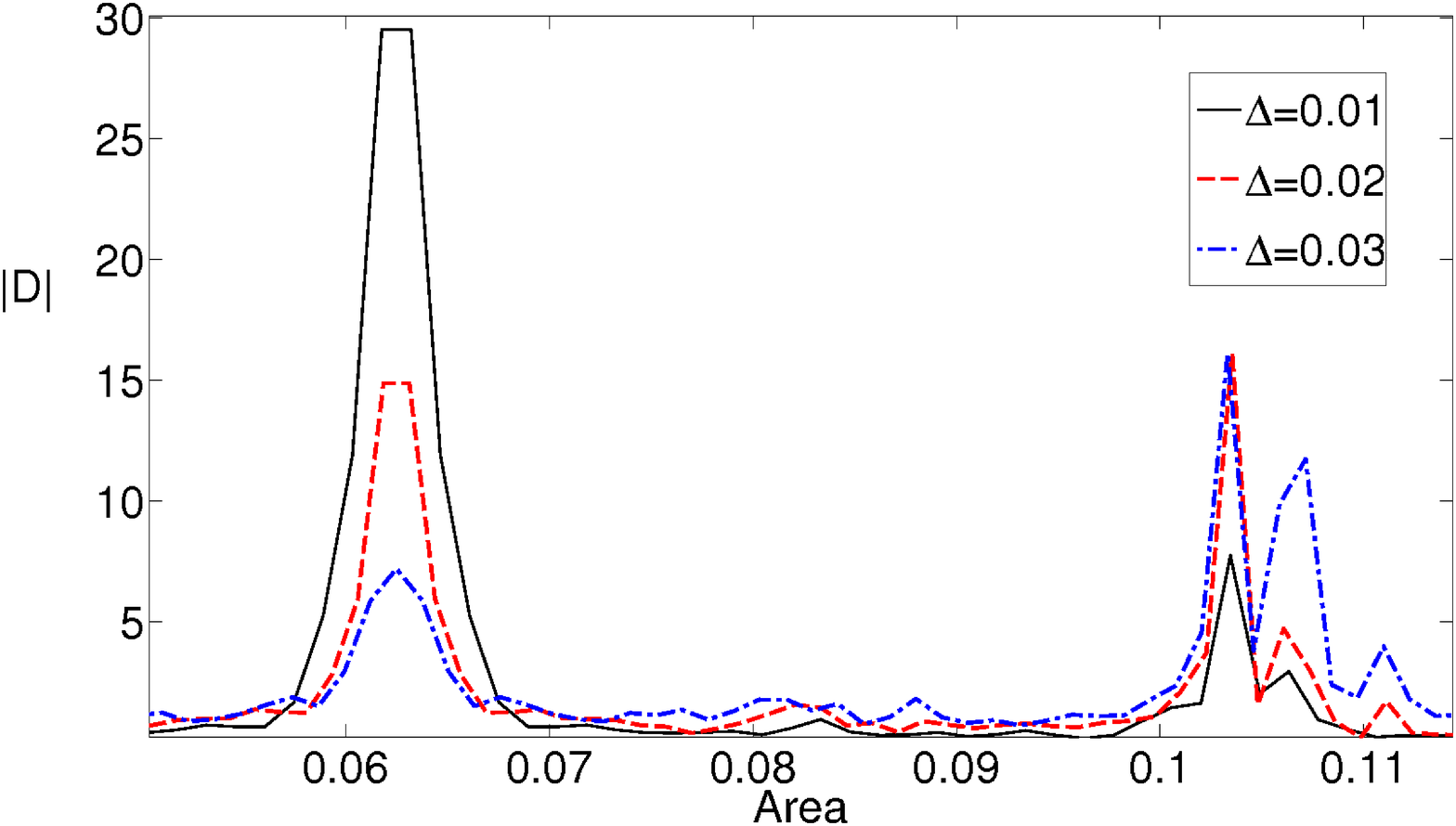}
\caption{Power spectrum for oscillations of the position of the lowest LL for small values of the pairing potential amplitude, $\Delta$. As discussed in the text, the peak at 0.0625 corresponds to the area $A_T$, the original FS. The peak at 0.103 corresponds to $A_T-A_b$, the boomerang-shaped area shown in Fig.~\ref{boomerang}a, while the feature at 0.1065 corresponds to the orbit with area $2A_b-A_T$, shown in Fig.~\ref{boomerang}c.}
\label{varydelta}
\end{figure}

However, once again, this is not the whole story.  We should look for oscillations in the position spectrum due to the orbit shown in Fig.~\ref{link}c, involving four Andreev-Bragg scatterings, which is shown for the period 8 system in Fig.~\ref{boomerang}c.  This feature is expected to be weak for $\Delta = 0.02$ and to occur at $2A_b-A_T = 0.2315$. Subtracting 1/8, we expect a small peak in the position spectrum at 0.1065, which is barely visible in Fig.~\ref{power}.   In order to check whether this feature is real or just an artifact, we vary the value of $\Delta$. The results are shown in Fig.~\ref{varydelta} for $\Delta$ = 0.01, 0.02, and 0.03.  As expected, the magnetic breakdown peak at 0.0625 drops precipitously with increasing $\Delta$ while the ''boomerang" peak at 0.103 grows and the peak at 0.1065 due to the closed orbit grows more rapidly.

\section{Intermediate and large $\Delta$}

\begin{figure} [tbph] 
\includegraphics[scale=.2]{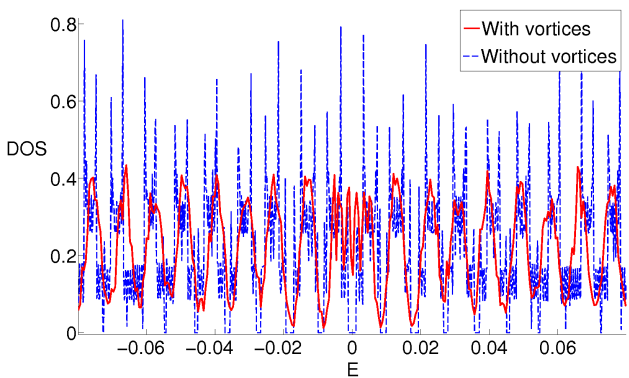}
\caption{Comparison of the low-energy DOS of a $\pi$-striped superconductor in the presence of a magnetic field of $L=1024$ with $\Delta=0.25$ and $\mu=-0.3$ corresponding to $1/8$ doping with and without vortices.} 
\label{mu3comp}
\end{figure}

In this section, we consider larger values of $\Delta$, specifically the range $0.15\le\Delta\le 0.6$. For this range, the FS is rather different from the case when $\Delta$ is very small.  The difference is illustrated in the right-hand panels of  Fig.~\ref{sp02and25}.  When $\Delta$ is small, the FS of Fig.~\ref{sp02and25}a, constructed by repeated translations of the FS of the first BZ, consists of overlapping shapes of the type shown in Fig.~\ref{boomerang}c. On the other hand, for $\Delta = 0.25$, the FS in the right hand panel of Fig.~\ref{sp02and25}b consists, to a first approximation, of interwoven open orbits, four each for positive and negative values of $k_y$.  In fact, although it is difficult to see in Fig.~\ref{sp02and25}b, there are small gaps in these FS sections wherever two of them cross.  These gaps are vanishingly small for $\Delta \approx 0.15$ and increase with increasing $\Delta$.  For larger values of $\Delta$, $0.4\le\Delta\le 0.6$, the FS sections resemble rows of hour-glass-shaped figures as will be shown below.  We note that, for these larger value of $\Delta$, closed orbits result from Andreev-Bragg scattering at, and tunnelling across, the small FS gaps and, as we shall see, lead to quantum oscillations.  These oscillations all correspond to  areas less than $1/8$ of the BZ, and hence the fact that the FS areas that we calculate by the semiclassical method are only defined modulo $1/8$ of the BZ is not important when $\Delta$ is large.

Before proceeding further, we verify that the method works, at least for intermediate $\Delta$, by comparing results for the density of states of the semiclassical case to that of the exact BdG method with vortices. It is found that the two cases are in qualitative agreement, as shown in Fig. \ref{mu3comp} for the low-energy DOS for $\Delta=0.25$ and $\mu=-0.3$ corresponding to $1/8$ doping. This reinforces the earlier comparison of the energy bands along $Y \rightarrow \Gamma \rightarrow X$, for these parameters, which was shown in Fig. ~\ref{bands}. Note that the nonzero DOS at $E=0$ for the case with vortices is a commensurability effect which is absent for the case without vortices. Also note that agreement between calculations with and without vortices is not found for large values of $\Delta > 0.35$, since it was shown in Fig.~(9) of our earlier work\cite{zelli} that LL are not found for the exact BdG calculation in that regime.  

Next we study how quantum oscillations behave for the cases of intermediate and large $\Delta$  within the semiclassical approximation. As in the small $\Delta$ case, one can measure the width and position of the peak closest to the Fermi energy. Here, the results are discussed in two subsections, at half-filling and around $1\over 8$ doping.  We consider a large range of $\Delta$, including where the semiclassical approximation is not valid, because this analysis is helpful for identifying which Fermi surface trajectories are responsible for the observed frequencies, and it can be extrapolated to intermediate $\Delta$ where the approximation is reliable.

\subsection{Half-filling}

\begin{figure} [tbph]
\includegraphics[scale=.2]{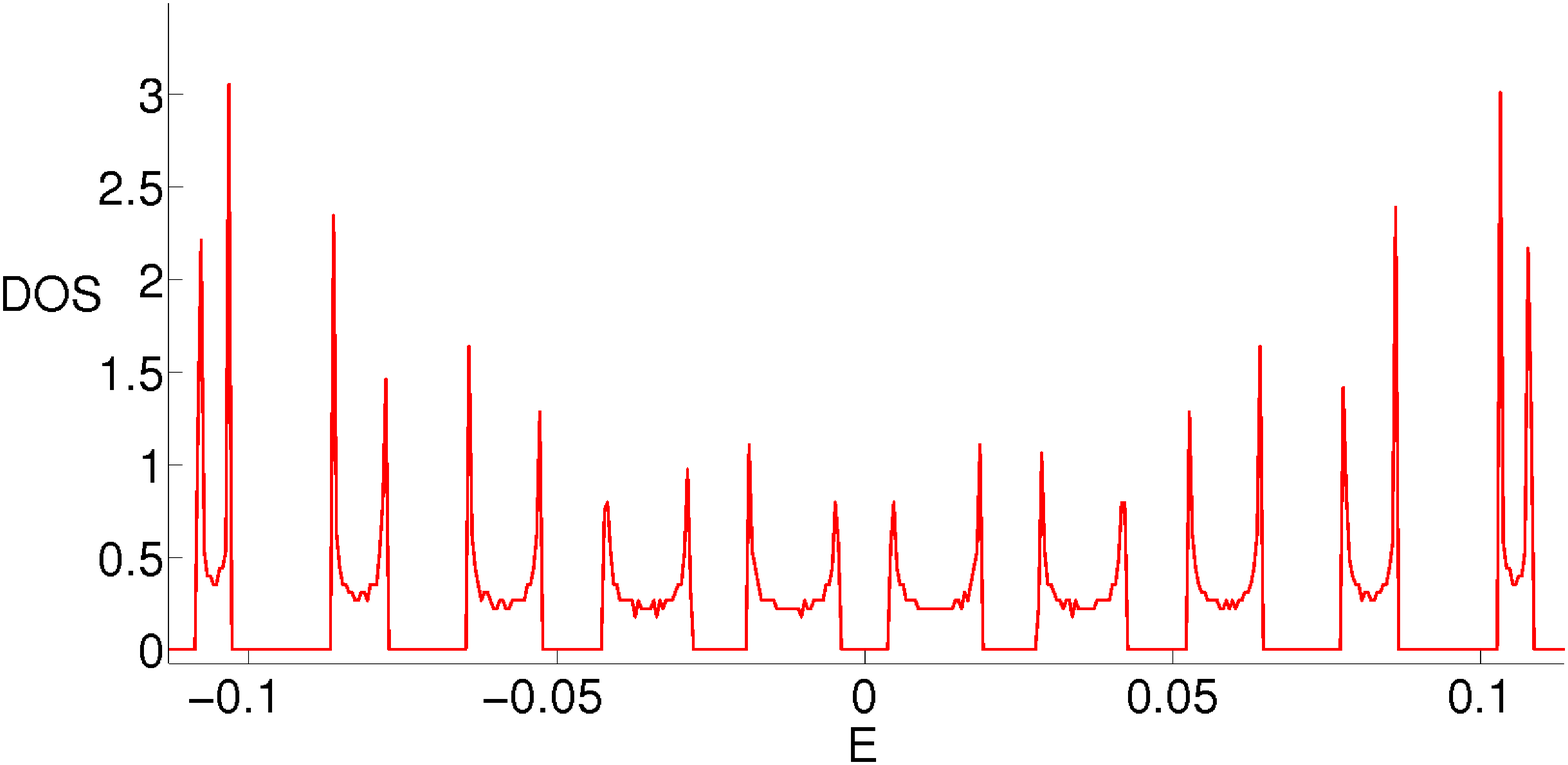}
\caption{The low-energy DOS for $\Delta=0.4$ and $L=800$ at half-filling. Each (double) peak has twice the degeneracy of a LL.} 
\label{dos2mu0}
\end{figure}

For $\mu=0$, the points at the centers of the Fermi arcs, which occur at $k_y = \pm \pi/2$, are gapless. In addition, for this special case of $\mu =0$, the FS arcs for $k_{y}> 0$ ($k_{y}<  0$) are symmetric under reflection across the line $k_{y} = \pi/2$ ($k_{y} = -\pi/2$).  
 
Fig.~\ref{dos2mu0} shows the low-energy DOS for $\Delta=0.4$ at half-filling in the presence of a magnetic field of $L=800$. Each peak has twice the degeneracy of a LL and is, in fact, composed of two Landau levels that touch.  To see this, it is only necessary to turn on a small negative chemical potential which creates a small gap at the center of the peak. This merging of pairs of Landau levels does not occur in the case with vortices, where the Landau levels are resolved even at half-filling. .

\begin{figure} [tbph]
\includegraphics[scale=.2]{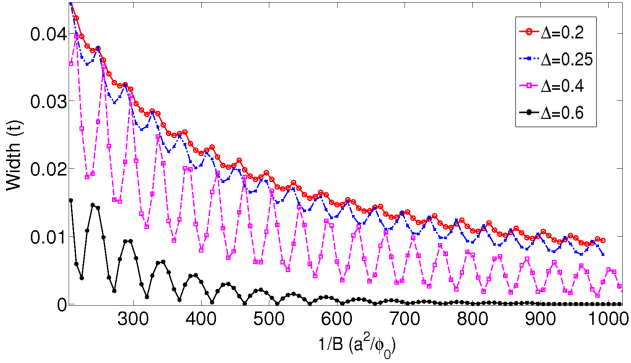}
\caption{Half width of the peak closest to $E=0$ for different values of $\Delta$ at half-filling. The Fermi surfaces for two of the $\Delta$ values in this figure are shown in Fig.~\ref{fsmb}.} 
\label{wdifdel}
\end{figure}

Fig.~\ref{wdifdel} shows the width of the first peak as a function of $1/B$ for several values of $\Delta$ at half-filling.   The two most conspicuous features of this figure are a smooth background which decreases for decreasing $B$ and increasing$\Delta$ and oscillations which become more prominent for larger $\Delta$ and whose amplitude tends to decrease for decreasing $B$.

\begin{figure} [tbph] 
\includegraphics[scale=.233]{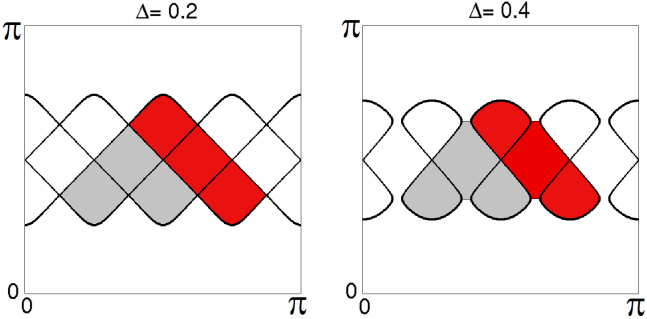}
\caption{Areas consistent with the quantum oscillations seen in the width of the first peak in the low-energy DOS are shown in red (dark-shaded) for two values of $\Delta$ at half-filling. Note that, for $\mu = 0$ the gray (light-shaded) areas have the same area as the red areas.}
\label{fsmb}
\end{figure}

The behavior of Fig.~\ref{wdifdel} can be understood by comparing the left and right panels of Fig.~\ref{fsmb}.  The right hand panel, for $\Delta = 0.4$, shows a line of figure-eight-shaped Fermi surfaces which are separated by gaps in $k$-space, in contrast to the left hand panel, for $\Delta = 0.2$, which appears to show a set of four interwoven open orbits.  Closer scrutiny shows that the apparently continuous lines in the left hand panel have small gaps at avoided crossings.  At high fields, magnetic breakdown causes tunnelling across these gaps along the open orbits.  Alternatively, four successive Andreev-Bragg reflections give rise to the figure-eight orbits which enclose zero net flux for $\mu = 0$ because the two identical lobes are traversed in opposite directions.   Motion along open orbits and figure-eights contributes to the smooth background for the widths shown in Fig.~\ref{wdifdel}.  Quantum oscillations occur when Andreev-Bragg scattering at the gaps leads to closed orbits.  Closed orbits involving two Andreev-Bragg scatterings and two tunnellings are shown by the red (dark-shaded) areas in Fig.~\ref{fsmb}.

\begin{figure} [tbph] 
\includegraphics[scale=.2]{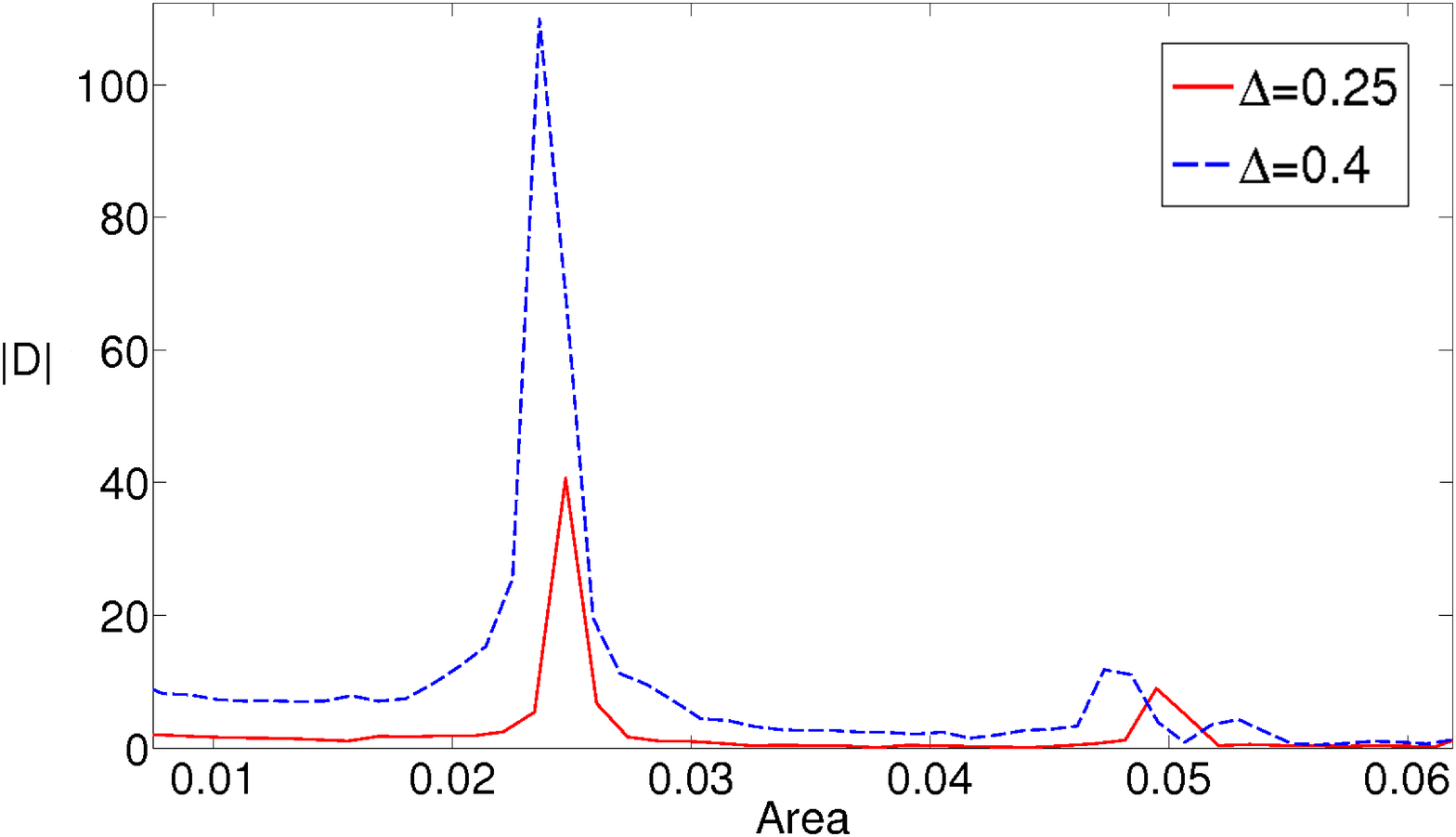}
\caption{Power spectrum associated with the oscillations in the width for $\Delta=0.25$ and $\Delta=0.4$ at half-filling. The $x$ axis is rescaled so that it corresponds to area in units of the area of BZ.} 
\label{m125area}
\end{figure}

\begin{figure} [tbph] 
\includegraphics[scale=.2]{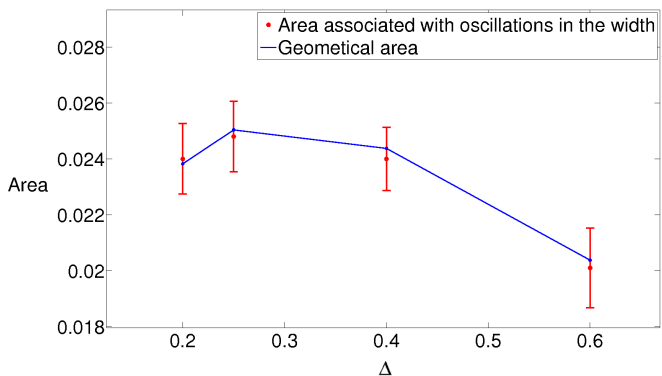}
\caption{Comparison of the geometrical area (red or gray area in Fig. \ref{fsmb}) and the area associated with quantum oscillations in the width of the lowest energy peak for different values of $\Delta$ at half-filling.} 
\label{areavsdel}
\end{figure}

Fig.~\ref{m125area} shows the power spectrum associated with the oscillations in the width of the lowest energy peak for $\Delta=0.25$ at half-filling. A sharp peak appears in this spectrum around $0.025$, along with a second one that seems to correspond to a second harmonic. The area associated with quantum oscillations for other $\Delta$ values in Fig. \ref{wdifdel} are also calculated and are found to be consistent with the red colored (dark-shaded) areas shown in Fig. \ref{fsmb}. Note that the gray (light-shaded) areas have the same area as the red (dark-shaded) areas. This is because, at half-filling, the two loops in the figure-eight segments have the exact same area. The consistency is shown in Fig. \ref{areavsdel} where, for different $\Delta$, we compare the geometrical area corresponding to the red (or gray) regions in Fig.~\ref{fsmb} to the area associated with quantum oscillations. 

It is worth noting that the average position of the lowest energy peak (which consists of two LLs) does not exhibit quantum oscillations, but rather scales linearly with $B$ as expected for Landau levels.  This is because the two Landau levels in this peak oscillate in opposite directions.  As a result, the oscillations in the width of this feature also reflect position oscillations of its two components. 

When $\Delta$ is very large, as in the lowest curve of Fig. ~\ref{wdifdel}, magnetic breakdown is suppressed, and the low-energy LL features are very sharp. 

\begin{figure} [tbph] 
\includegraphics[scale=.2]{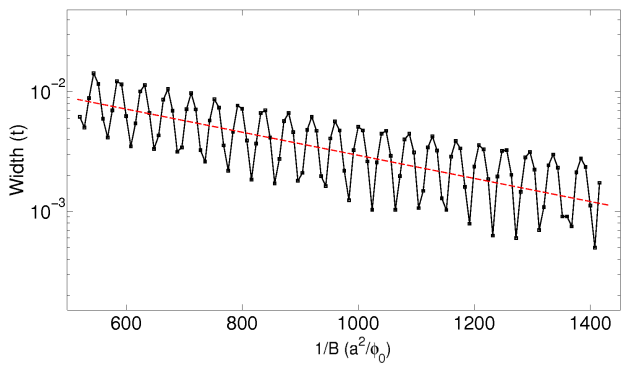}
\caption{Semi-log plot of the width of the first LL for $\Delta=0.4$ at half-filling as a function of $1/B$ showing a fairly linear average behavior for not very large fields. This is expected if the broadening is caused by magnetic breakdown. The dashed line is a linear fit to the data.} 
\label{log}
\end{figure}

To summarize so far, we have seen that, at half-filling, sharp peaks with the degeneracy of two Landau levels are formed for very large $\Delta$ where the figure-eight-shaped FS segments are well-separated. As $\Delta$ decreases, the gaps between figure-eight segments decrease and magnetic breakdown occurs which leads to broadening of the peaks. This is reflected in the smooth non-oscillatory part of the curves in Fig. \ref{wdifdel}. According to the theory of magnetic breakdown,\cite{shoenberg} its probability is proportional to $\exp(-B_{0}/B)$ where $B_{0}$ is a constant. Taking the broadening of the first peak as an estimate of the probability of magnetic breakdown, we show the width as a function of $1/B$ in semi-logarithmic plot for $\Delta=0.4$ in Fig. \ref{log}. The non-oscillatory part exhibits a linear behavior in this semi-log plot which further supports our argument that magnetic breakdown is responsible for broadening of the Landau levels.

\subsection{Nonzero $\mu$}
Away from half filling, for example at $1\over 8$ doping, the Landau levels are well resolved. Each peak has a number of states close to that of a LL, and the total number of states in peaks that are related by $E \to -E$ is exactly twice the degeneracy of a LL.  This behavior is consistent with BdG calculations with vortices, as shown in Fig.~\ref{mu3comp}.

\begin{figure} [tbph] 
\includegraphics[scale=.21]{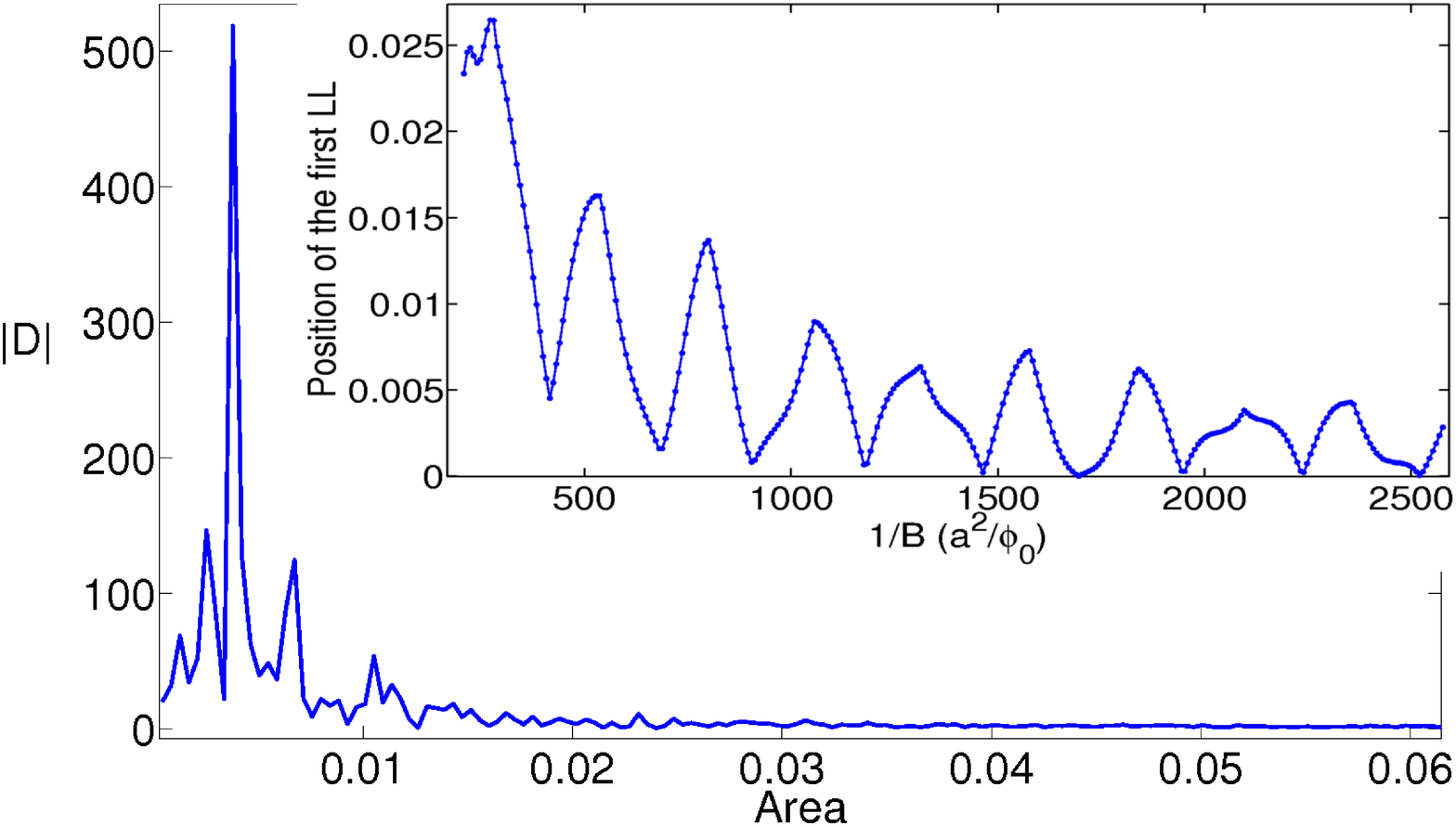}
\caption{Power spectrum associated with the position of the first LL for $\Delta=0.6$ and $\mu=-0.5$. The inset shows the position of the first LL for the same parameters.} 
\label{d3m5power}
\end{figure}
To better understand the quantum oscillations that exist in a $\pi$-striped superconductor, we start from the very large $\Delta$ limit where the Landau levels are sharp and magnetic breakdown is strongly suppressed. 

\begin{figure} [tbph] 
\includegraphics[scale=.21]{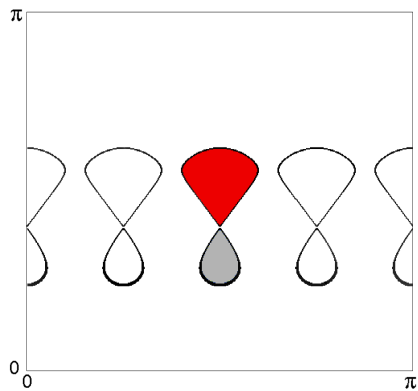}
\caption{FS for $\Delta=0.6$ and $\mu=-0.5$. The difference in the area of the the gray (light-shaded) and red (dark-shaded) areas gives rise to the strongest peak in the power spectrum of the position of the first LL.} 
\label{sp6}
\end{figure}
The position of the first LL for $\Delta=0.6$ and $\mu=-0.5$ is plotted in the inset of Fig. \ref{d3m5power} as a function of magnetic field. The position shows an oscillatory behavior with a long period, which implies that the QO area is small. The power spectrum associated with the position of the first LL for $\Delta=0.6$ and $\mu=-0.5$ is shown in Fig. \ref{d3m5power}. Within error bars, the largest peak corresponds to the {\em difference} in the areas of the gray (light-shaded) and red (dark-shaded) areas shown in Fig. \ref{sp6}, which are traversed in opposite directions. The other two peaks on either side of the main peak correspond to the separate gray (light-shaded) and red (dark-shaded) areas. These arise due to a small gap where the two lobes meet, leading to small amplitude reflections into closed orbits around each lobe.  Except for these small peaks, the oscillatory behavior that we measure corresponds predominantly to orbits around the figure-eight-shaped areas.  For this value of $\Delta$, there is no sign of magnetic breakdown across gaps separating neighboring figure-eights.

\begin{figure} [tbph] 
\includegraphics[scale=.21]{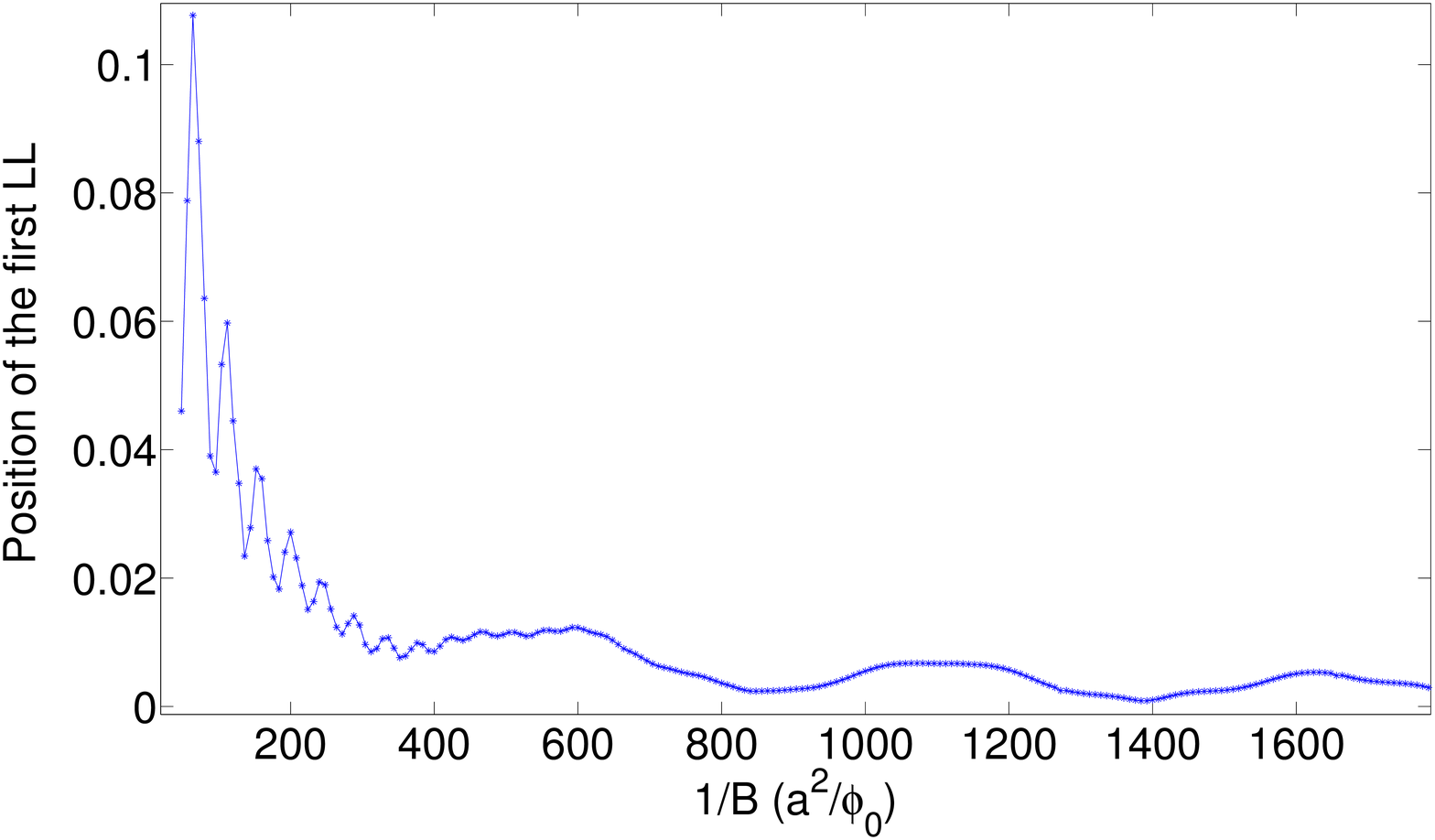}
\caption{Position of the first LL for $\Delta=0.5$ and $\mu=-0.4$.} 
\label{d25m4p}
\end{figure}
\begin{figure} [tbph] 
\includegraphics[scale=.22]{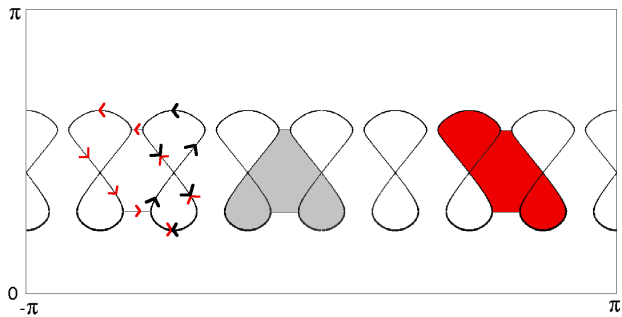}
\caption{FS for $\Delta=0.5$ and $\mu=-0.4$. The red area is associated with short-period oscillations in Fig. \ref{d25m4p} for larger magnetic fields and the gray (light-shaded) area is associated with the oscillations in the width of the first LL when magnetic breakdown occurs. The difference in the area of the two lobes of the figure-eight results to long-period oscillations in Fig. \ref{d25m4p} at smaller fields. Black (thin) and red (thin) arrows show the two possible semiclassical paths.} 
\label{sp5}
\end{figure}

Now we decrease $\Delta$ by a small amount in order to see what happens when magnetic breakdown is possible. Fig. \ref{d25m4p} shows the position of the first LL for $\Delta=0.5$ and $\mu=-0.4$. For larger magnetic fields, the short-period oscillations are due to magnetic breakdown and correspond to the red (dark-shaded) area shown in Fig. \ref{sp5}. Magnetic breakdown does not occur for smaller magnetic fields, and so only long-period oscillations occur at small $B$, corresponding to the difference in the areas of the two lobes in the figure-eight-shaped areas of Fig. \ref{sp5}. 

This provides the key to understanding the semiclassical motion. One possible semiclassical motion is shown by the black arrows in Fig. \ref{sp5}. The phase that a quasiparticle gains by going around this path is proportional to the difference in the areas of the two lobes of figure-eight. The semiclassical motion associated with magnetic breakdown is shown by the red (thin) arrows. In this case, the phase gained by precessing around the path is proportional to the red (dark-shaded) area. Like the small $\Delta$ case, we expect that the difference of the two paths to determine oscillations in the width of the position peak. Indeed this is what happens. The area associated with the oscillations in the width is equal to the gray (light-shaded) area in Fig. \ref{sp5}. 

\begin{figure} [tbph] 
\includegraphics[scale=.21]{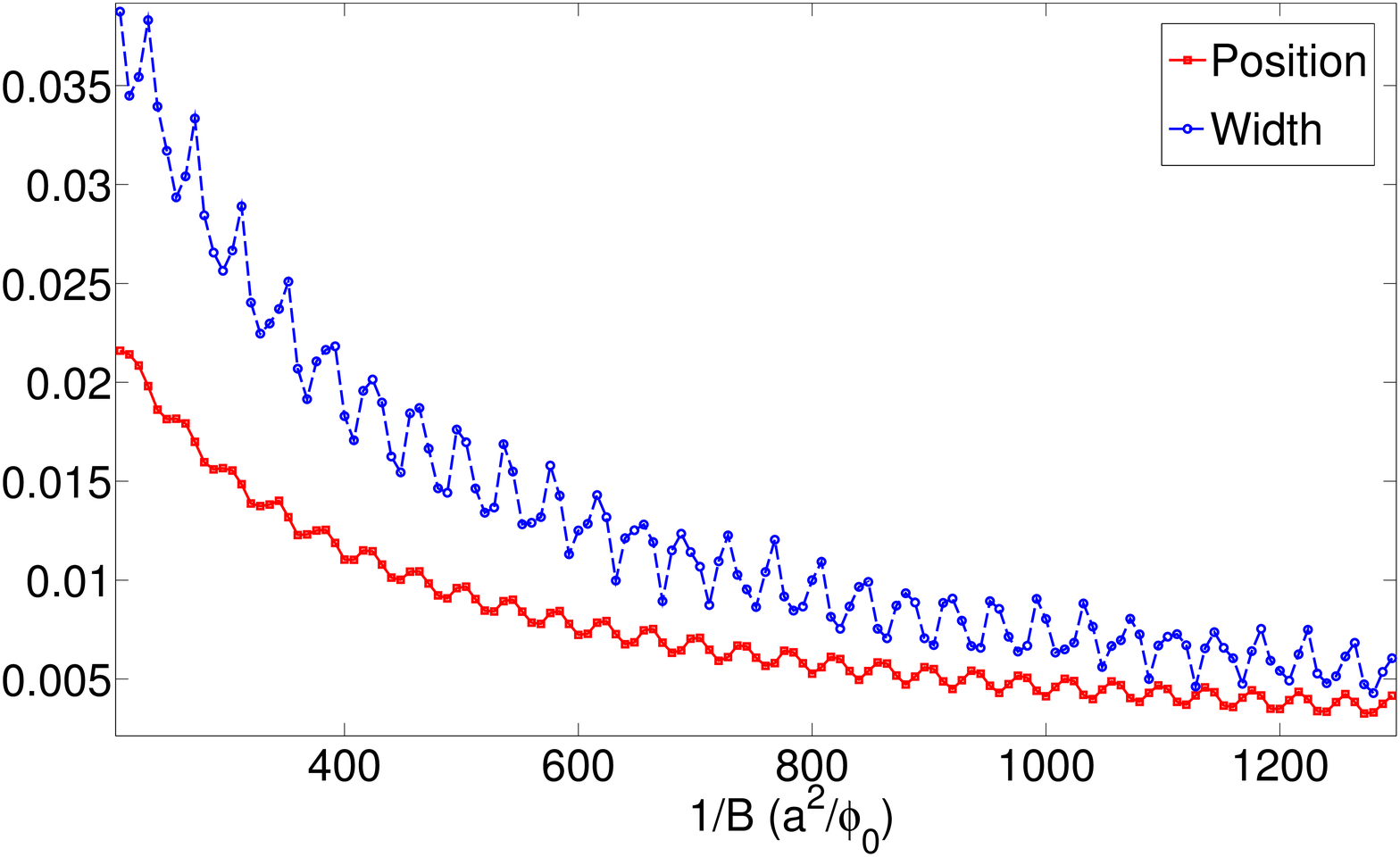}
\caption{Position and width of the first peak for $\Delta=0.25$ and $\mu=-0.3$, corresponding to $1\over 8$ doping, plotted versus $1/B$.} 
\label{wp}
\end{figure}

Having gained some physical insight from the case of very large $\Delta$, we move on to the case of smaller $\Delta$. In Fig. \ref{wp}, we show the width and position of the first LL for $\Delta=0.25$ and $\mu=-0.3$ corresponding to $1\over 8$ doping. Both quantities show an oscillatory behavior as a function of $1/B$. The amplitude of oscillations is larger for the width and the frequency is slightly higher.

\begin{figure} [tbph] 
\includegraphics[scale=.21]{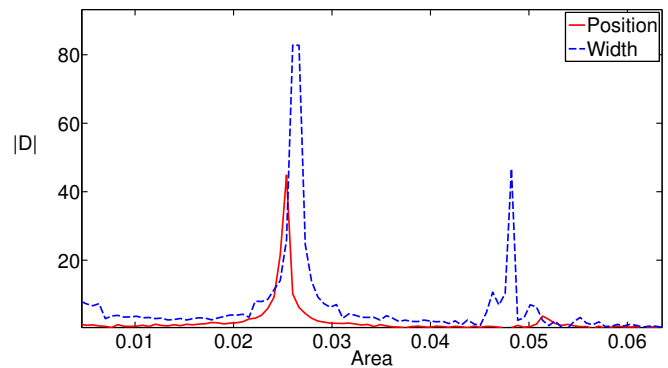}
\caption{Power spectrum for $\Delta=0.25$ and $\mu=-0.3$} 
\label{fulls}
\end{figure}
\begin{figure} [tbph] 
\includegraphics[scale=.21]{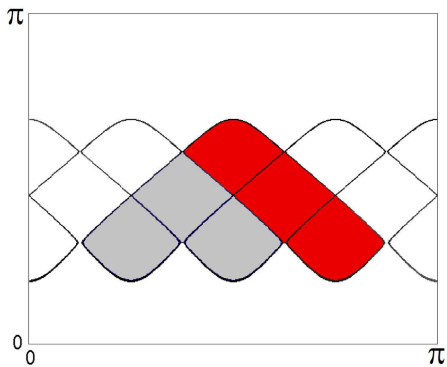}
\caption{FS for $\Delta=0.25$ and $\mu=-0.3$ in the quadrant of the first BZ.} 
\label{sp25}
\end{figure}

\begin{figure} [tbph] 
\includegraphics[scale=.2]{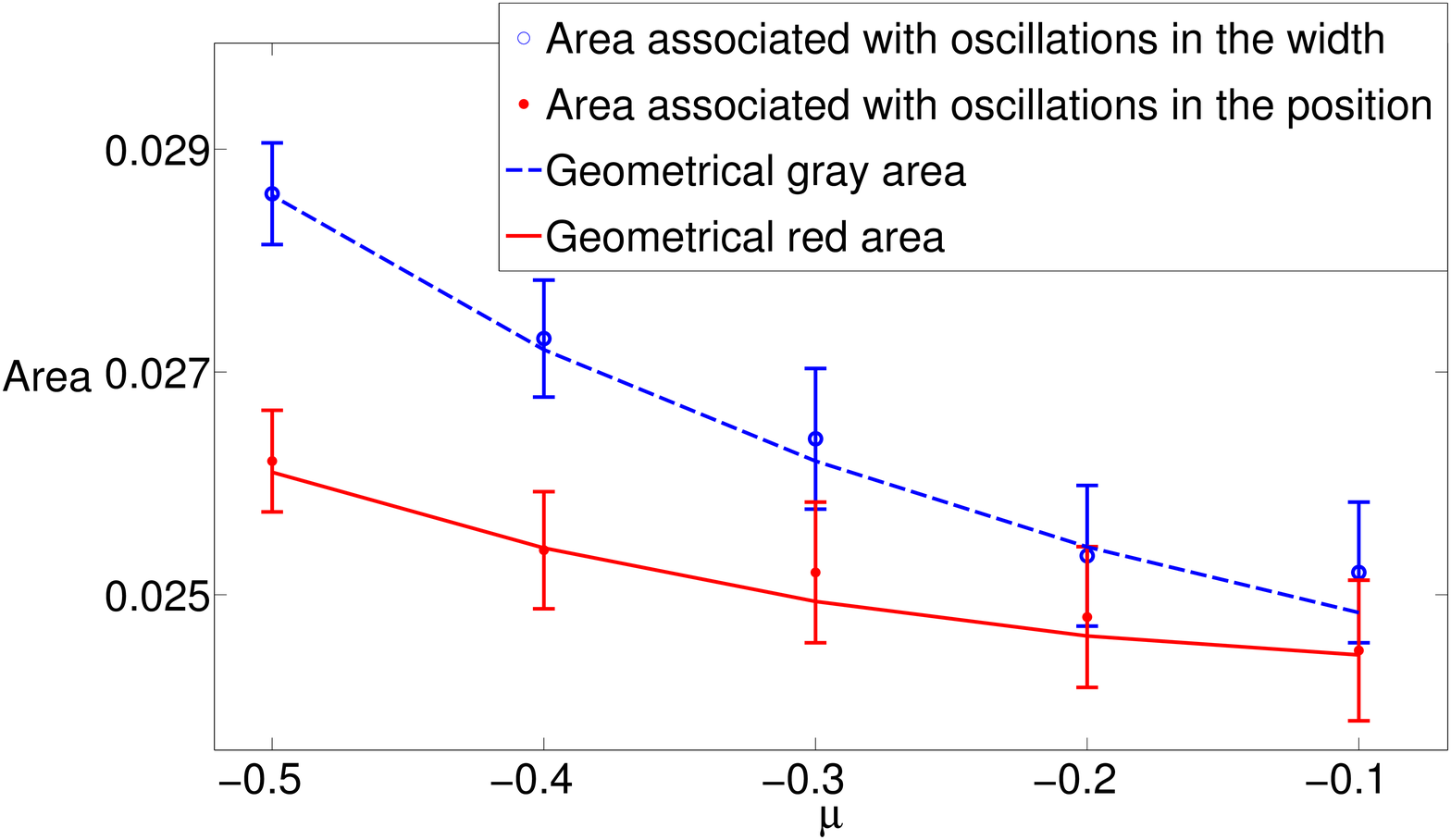}
\caption{Comparison of the geometrical area and the area associated with quantum oscillations in the width and position of the first LL as a function of $\mu$ for $\Delta=0.2$. The geometrical area is the area corresponding to the red (dark-shaded) region in Fig. \ref{sp25} in the case of $\Delta=0.2$.} 
\label{wparea}
\end{figure}

The power spectra associated with the position and width of the first LL for $\Delta=0.25$ and $\mu=-0.3$, corresponding to $1\over 8$ doping, are shown in Fig. \ref{fulls}. For simplicity, we limit our discussion to the largest position and width peaks which lie between 0.02 and 0.03 of the BZ. The position spectrum exhibits a peak at around $0.025$ which is due to magnetic breakdown and is associated with the red (dark-shaded) area in Fig. \ref{sp25}. In the width spectrum, there are two peaks. The first one, which is larger, is associated with the gray (light-shaded) area shown in Fig. \ref{sp25}. Note that the gray area can be thought as the red area minus the difference in the areas of the two loops of the figure-eight. In Fig. \ref{wparea}, we have shown the consistency between the position and width spectra of the first peak and the geometrical area for $\Delta=0.2$ as a function of the chemical potential. As $\mu$ becomes more negative, the area associated with the width oscillations becomes larger than the area associated with the position oscillations. This is consistent with the fact that the area of the lower loop of the figure-eight segments is larger than the upper loop for this smaller value of $\Delta$. We will see in the next section that, near $1\over 8$ doping, the period of the oscillations in the specific heat, as calculated for this model, corresponds to that seen for the position of the first LL.

\begin{figure} [tbph] 
\includegraphics[scale=.2]{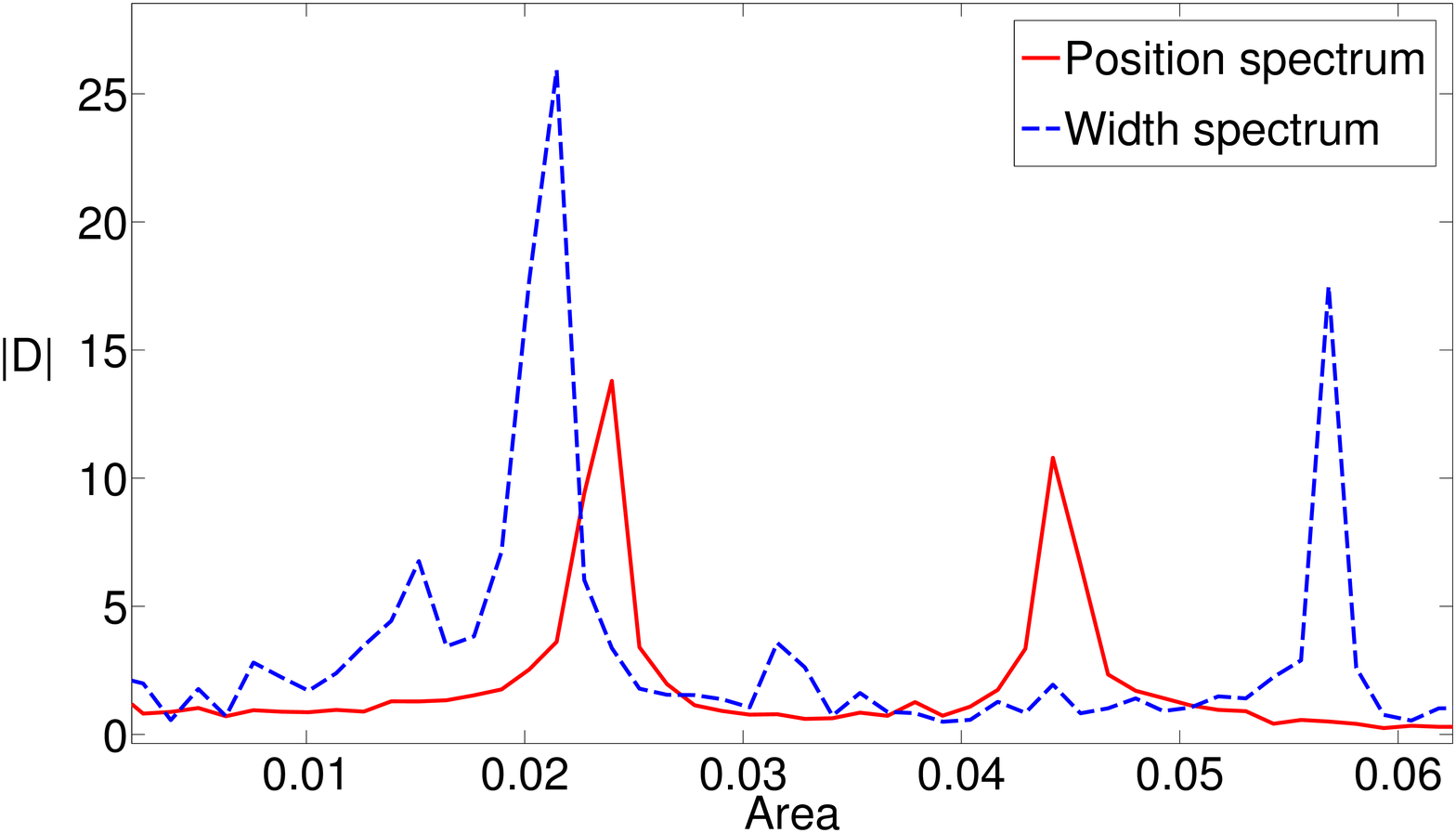}
\caption{The spectra associated with oscillations in the width and position for $\Delta=0.25$ and $t_{2}=-0.15$ at $1\over 8$ doping. The peaks correspond to the gray (light-shaded) and red areas shown in Fig.~\ref{t2area}. The results are consistent with those for $t_{2}=0$.}
\label{t2}
\end{figure}

\begin{figure} [tbph] 
\includegraphics[scale=.21]{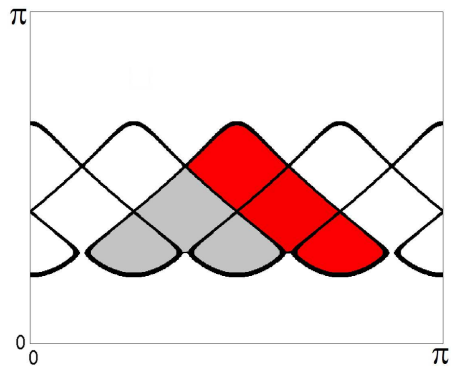}
\caption{The areas associated with the first peaks of the position and width spectra in Fig.~\ref{t2} for $\Delta=0.25$ and $t_{2}=-0.15$ at $1\over 8$ doping.}
\label{t2area}
\end{figure}

So far all the calculations were for the case where the second nearest neighbor hopping term was set to zero. To allow for the possibility of a more realistically shaped FS, calculations were also performed for $\Delta=0.25$ and $t_{2}=-0.15$ at $1\over 8$ doping. The results are as expected from the $t_{2}=0$ calculations. The power spectrum for oscillations in the width and position of the lowest LL are shown in Fig. \ref{t2}. The first peak associated with oscillations in the position of the first LL corresponds to the red (dark-shaded) area in Fig. \ref{t2area}. The first peak associated with the width of the first LL corresponds to the gray (light-shaded) area which is smaller than the red (dark-shaded) area. The calculation for non-zero $t_{2}$ demonstrates that the position and width frequencies are sensitive to the details of the band structure. Hence, the band structure could, in principle, be used to fit theory to experiment.

\section{Specific Heat}
\begin{figure} [tbph] 
\includegraphics[scale=.194]{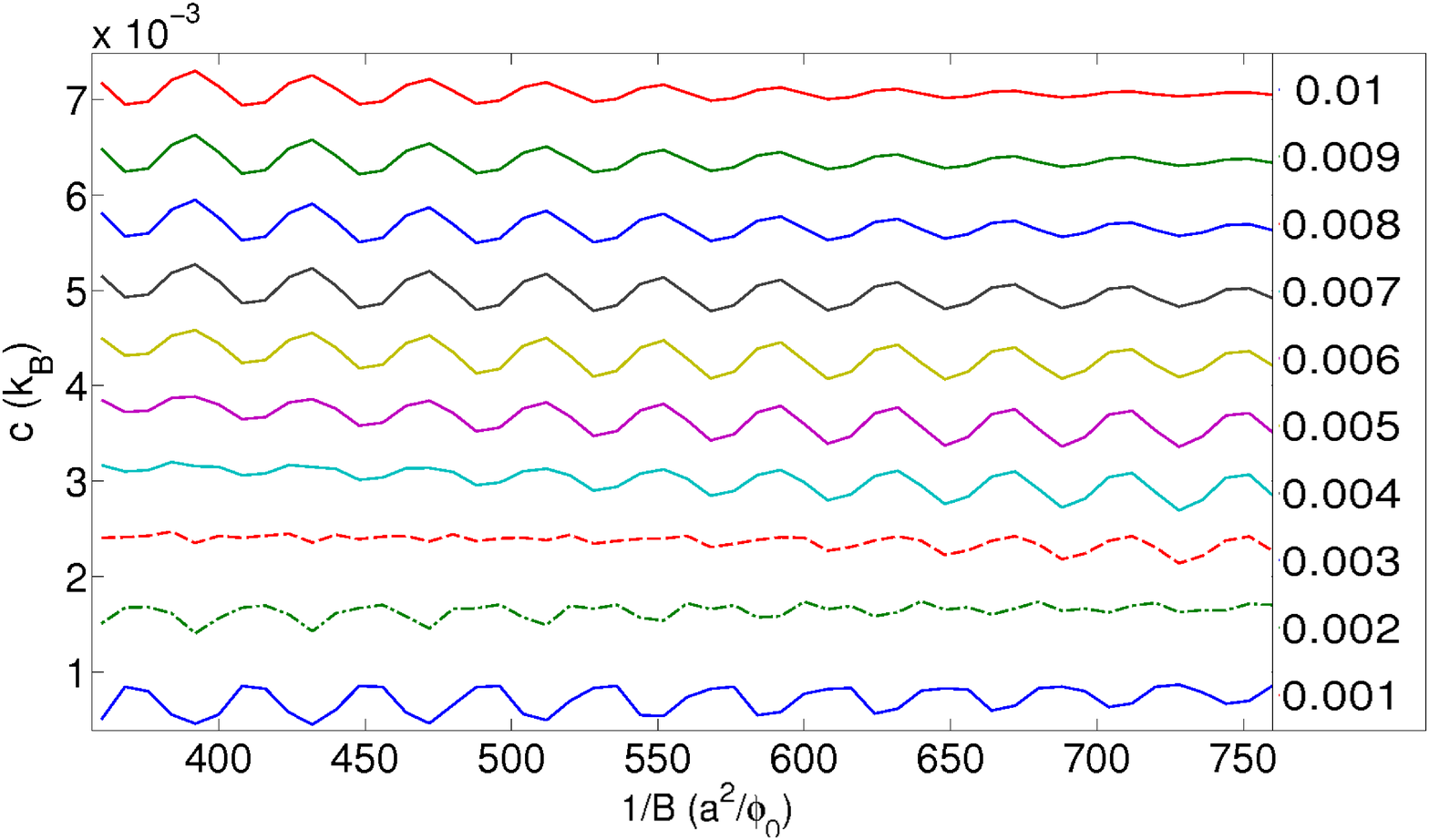}
\caption{Specific heat versus $1/B$ for $\Delta=0.25$ and $\mu=-0.3$ and $t_{2}=0$ for different temperatures. Temperatures in units of the hopping term, $t$, are shown on the right. Note the $\pi$ phase shift in the oscillatory behavior of specific heat as T increases through $T^{*}\approx 0.003t$.} 
\label{c25}
\end{figure}
The question remains whether oscillations, related to those seen in the width and the position of the first LL, can be observed in a physically measurable quantity. In this section, we calculate the specific heat in order to make a connection to experiment. Here, the same method, which involves a sum over all excited quasiparticle states, and assumptions are made as in our earlier work, Ref. \onlinecite{zelli}. In that paper, it was shown that the specific heat of the model could be made consistent with the observed specific heat of a cuprate superconductor at $1\over 8$ doping in zero field or in the presence of a magnetic field by adjusting the value of the only parameter in the model, $t$. (Note that in our earlier work and in this section we take $t_2 = 0$.)  In our earlier work, the field dependence of the specific heat could not be studied in detail for the same reasons that quantum oscillations could not be measured, and, in addition, commensurability effects were exaggerated because of the restriction to commensurate vortex arrangements. Using the semiclassical approximation of this study, the magnetic field can be changed in relatively small steps, and, in addition,  commensurability effects are not present. As a result, we are able to observe quantum oscillations in the specific heat. 

Fig.~\ref{c25} shows the specific heat versus $1/B$ for $\Delta=0.25$ and $\mu=-0.3$ at different temperatures. The oscillatory behavior corresponds to the same area as seen in the position oscillations of the first peak in  Fig. \ref{fulls} and corresponds to the red area shown in Fig. \ref{sp25}. Interestingly, there is a $\pi$ shift in the oscillatory behavior of the specific heat at a temperature $T^{*}$. This is consistent with the Lifshitz-Kosevich (LK) formula for the specific heat.
\begin{figure} [tbph] 
\includegraphics[width=3in]{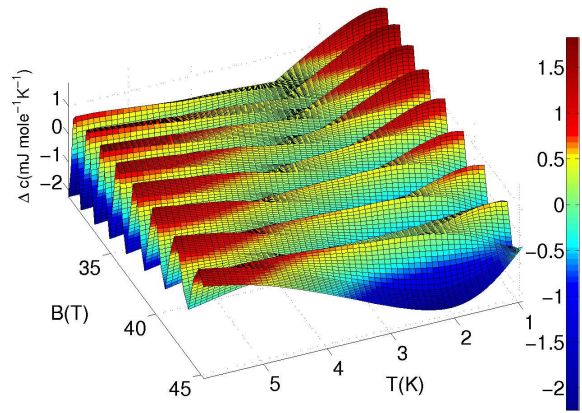}
\caption{The oscillatory part of the calculated specific heat for $\Delta=0.25$ and $\mu=-0.34$ with a zero second nearest neighbor hopping shown as a function of the magnetic field and temperature. To plot the data, $t=0.16$eV is chosen.} 
\label{3D}
\end{figure}
\begin{figure} [tbph] 
\includegraphics[scale=.2]{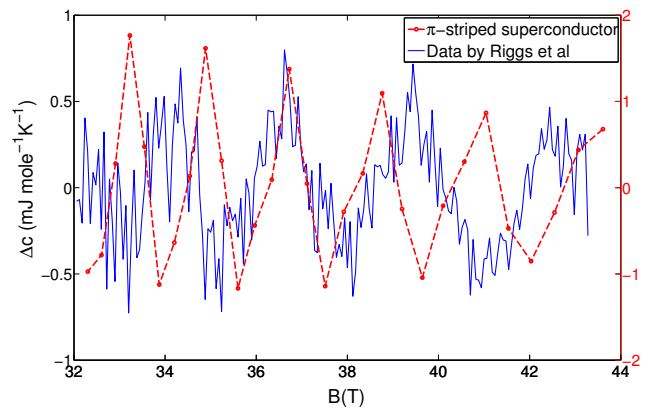}
\caption{The oscillatory parts of the specific heat data by Riggs et al. and calculations for a $\pi$-striped superconductor at $T=1$K. The left y-axis scale is for the experimental data and the right one is for the model.} 
\label{x1K}
\end{figure}

To make a direct connection to the experimental data by Riggs et al., we have shown the oscillatory part of our specific heat calculations for $t=0.16$eV in Fig. \ref{3D}. The figure can be compared to Fig. 2a of Ref. \onlinecite{riggs}. The qualitative agreement is good bearing in mind that we have used only one parameter $t$ to fit the data. In addition, we have compared the oscillatory behavior part of the data in our model to the experimental data at $T=1$K in Fig. \ref{x1K}. The period of oscillations is shorter for our data because the quantum oscillations area is larger by about $20\%$. The fact that the approximate magnitudes of the oscillations in the specific heat for the two data sets are similar supports the conjecture that the $\pi$-striped superconductor is a possible candidate model for explaining quantum oscillations in high $T_{c}$ cuprates. 

\section{Discussion and Conclusions}
In this paper, we have considered a model of sinusoidally modulated d-wave superconductivity, the $\pi$-stripe phase, in the presence of magnetic fields, and we have developed an approximate semiclassical method to calculate physical properties of this model as a nearly continuous function of field.  The model is distinctly different from conventional models of quantum oscillations in metals because of the paired nature of the quasiparticle states near the Fermi energy. In this model, the reconstructed FS arises from Andreev scattering by the periodic pairing potential.  In the presence of a magnetic field, the semiclassical approximation suggests that electrons and holes precess along reconstructed FS orbits and also tunnel between these orbits via magnetic breakdown. 

It is indeed surprising that LLs and quantum oscillations would exist in such a model, since the pairing mixes particles and holes and the charge of the quasiparticle excitations is not quantized.  However the numerical evidence for the existence of broadened LLs is compelling.  Furthermore, Andreev scattering appears to play a key role in the formation of the broadened LLs, since they do not exist for a charge density wave with an almost identical Fermi surface except for the absence of small gaps (of order $\Delta^3$) induced by Andreev scattering.
We attribute the fact that LLs exist to the fact that large regions of the Fermi surface remain sharp, even in the presence of the periodic pairing potential, with the quaiparticle charge essentially quantized as $e$ or $-e$.

Evidence for the existence of quantum oscillations is somewhat less convincing because of the necessity of neglecting superfluid flow effects (i.e., vortices) in order to do the calculations. It is reasonable to question how accurate it is to neglect effects due to the superfluid velocity on the energies and wavefunctions of the quasiparticles.  We have argued that a direct comparison of the low energy bands and the densities of states with and without vortices, for small and intermediate values of the gap amplitude, shows that, for both cases, these states resemble broadened Landau levels.  The effect of vortices is to modestly further broaden and distort the bands, but the result looks nothing like the case of uniform d-wave in a magnetic field, where vortex lattice and magnetic field effects are comparable.

      If superfluid flow and vortex lattice effects are indeed
      negligible, then
      one might expect that one could do calculations with vortices,
      changing the vortex lattice spacing in a way that, although
      unphysical, allows one to vary the magnetic field nearly
      continuously.  As examples of this, we have performed
      calculations where the vortex lattice spacing along $x$ is held
      fixed at 8 or 16, while the vortex spacing along $y$ is varied,
      one lattice spacing at a time, over a wide range.  For a lattice
      spacing of 8 along $x$, one can reproduce all the field values
      used in our calculations without vortices, while, for a lattice
      spacing of 16 along $x$ one can generate half those values. In all cases, broadened LLs are observed,
      with the commensuration effects mentioned previously which
      complicate
      analysis of the lowest LL position and width. However, we have
      calculated the low temperature specific heat and we do
      not observe
      the quantum oscillations that were found in the
      semiclassical calculations without vortices.

       We would note that these vortex lattice
      configurations and the circular superflow patterns are very
      artificial.   In reality
      the vortex lattice and the resulting superflow will adjust
      to conform to the modulated pairing potential for a given
      density of
      vortices.  
      We expect this relaxation to alter the commensuration effect and
      to reduce the perturbing effect of the superfluid
      velocity field.  Hence, we do not view the absence of 
      quantum oscillations for these configurations as evidence that
      quantum
      oscillations do not occur in the presence of an equilibrium
      vortex state. 
      Furthermore, quantum oscillations are observed
      in the cuprates under conditions where the state is
      resistive, i.e., a vortex liquid state rather than an ordered
      vortex lattice. Studying the properties of a $\pi$-modulated
      vortex liquid state is a challenging problem, but, again,
      one would expect the effects of the superfluid velocity to
      be small in such a state.

Another question which immediately comes to mind is whether such a state is likely to occur in nature or, more specifically, in the high $T_{c}$ cuprates.  Arguments for the occurrence of such a $\pi$-striped superconducting state have been given earlier by Berg, Fradkin and Kivelson.\cite{berg,stripe} Such states have also been studied by Baruch and Orgad.\cite{shirit}. In addition, there have been several numerical studies \cite{corboz,raczkowski,loder,himeda} of striped states that arise from the t-J model which find that the two states, one in which the gap oscillates in magnitude but does not change sign and the other in which the sign of the gap oscillates, are extremely close in energy.  One might expect that, in zero field, the nodeless state should win out, but the situation is likely to be different in non-zero field, where the $\pi$-stripe phase may have a lower Gibbs free energy.  If, in fact, the $\pi$-stripe phase is stabilized by a magnetic field, then the calculations in this paper would be directly relevant to observations of quantum oscillations in the cuprates. One could address the question of the relative stability of the $\pi$-stripe and nodeless stripe phases through self-consistent BdG calculations.  This requires having a microscopic Hamiltonian that stabilizes stripes at the mean field level.  Such calculations are left for future work.

At a more general level the $\pi$-stripe phase may be viewed as a type of FFLO state, where the mechanism is the underlying microscopic Hamiltonian, e.g. the t-J model, rather than Zeeman-splitting of the bands, and the gap modulation is microscopic and commensurate, rather than mesoscopic.  The phenomena which arise from the theory, a non-zero density of particle-hole states at the Fermi energy, the existence of Landau levels in a magnetic field, and the occurrence of quantum oscillations and magnetic breakdown are generic.  In particular, they do not depend on the superconductivity being d-wave. What is distinctive about such phases is that the frequencies of quantum oscillations will be different from those that arise from periodic modulation of the electron or spin density.  Of course one expects that, in general, these phenomena will coexist.  In particular, one expects that a sinusoidal modulation of the superconducting gap with wavevector {\bf Q} will induce modulations of the charge density with wavevector {2\bf Q}. 

Our method allows the calculation of quantum oscillations in physical properties, such as the specific heat presented in this paper, as well as oscillations in the magnetic susceptibility, resistivity and Hall resistivity which we have not yet attempted. For a reasonable model of the band structure, with nearest neighbor hopping and a modulated gap amplitude, $\Delta = 0.25t$, we find, near $1\over 8$ hole doping, a small frequency for the quantum oscillations which is similar to but slightly larger than what is observed experimentally. The calculated  temperature and field dependence of the specific heat are both similar to experiment.  For example, the phase of the specific heat oscillations reverses at a temperature $T^{*}(B)$ which can be well fit by setting the hopping parameter $t=0.16$eV.  Beyond this, it is difficult to make detailed comparison because our model is strictly two-dimensional and does not include disorder, and so the Dingle factor and the factor due to band warping are both unity.  One feature which is absent in this model is the background $\sqrt{B}$ dependence of the specific heat.  However, it is not clear from the data whether this $\sqrt{B}$ dependence persists to high magnetic field, or whether it is simply a low-field phenomenon.  The data of Riggs et al. could, in principle, correspond to a system which switches from a low-field d-wave superconductor to a high-field $\pi$-stripe phase. Whether such a transition would be sharp or broad depends on how sensitive it is to disorder and vortex liquid effects.

In conclusion, we have studied a system in which spatially modulated pairing induces a non-zero density of particle-hole states near $\rm E_F$ which, in the presence of a magnetic field, form broadened Landau levels and exhibit quantum oscillations. The nature of the reconstructed FS and the resulting orbits in a magnetic field are qualitatively different from that of a normal nearly-free electron metal. This type of behavior may occur in the high $T_{c}$ cuprates or possibly in other materials where superconductivity and stripe behavior coexist.

\begin{acknowledgements}
The authors would like to thank Scott Riggs for providing the data shown in Fig. \ref{x1K}.  We also benefitted from useful discussions with Steven Kivelson, Marcel Franz, Zlatko Tesanovic, Patrick Lee, K.-T. Chen, T. Senthil and Gilbert Lonzarich. This work was supported by the Natural Sciences and Engineering Research Council of Canada and the Canadian Institute for Advanced Research.
\end{acknowledgements}
\bibliography{bib}
\bibliographystyle{unsrt} 
\end{document}